\documentclass[journal, romanappendices]{IEEEtran}

\usepackage{times}
\usepackage{amsmath}
\usepackage{array}
\usepackage{url}
\usepackage{bm}
\usepackage{placeins}
\usepackage{float}
\usepackage{tabularx,colortbl}
\usepackage{amssymb}
\usepackage{booktabs}
\usepackage{xcolor}
\usepackage{graphicx}
\usepackage{cite}
\usepackage{subcaption} 

\providecommand {\abs}[1] {\, \lvert \, #1 \rvert \,}
\providecommand {\Abs} [1] { \left|  #1 \right | }
\providecommand {\pg} [1] { \left\{  #1 \right \} }
\providecommand {\pt} [1] { \left(  #1 \right) }
\providecommand {\pq} [1] { \left[  #1 \right] }
\providecommand {\fan} [1] { \phantom{#1} }
\providecommand{\dfr} [2] {\dfrac{#1}{#2} \,}
\providecommand{\fr} [2] {\frac{#1}{#2} \,}

\newcommand {\sqr} [1] { \sqrt{#1} \, }
\newcommand {\td} [1] { \tilde{#1} \, }
\newcommand {\re} [1] {\mathrm{Re} \pg{#1} }
\newcommand {\im} [1] { \mathrm{Im} \pg{#1} }

\newcommand {\uv}[1] {\mathbf{u}_{#1}}
\newcommand{\esp} [1] {\: \mathrm{e}^{#1} \,}
\newcommand {\jrm} {\mathrm{j}}
\newcommand {\dd} {\, \textrm{d}}
\newcommand {\ii} {\infty}
\newcommand {\bh}  {\mathrm{H}}
\newcommand {\bi}  {\mathrm{I}}
\newcommand {\bk}  {\mathrm{K}}
\newcommand {\pv}  {\mathrm{PV}}

\newcommand {\intii} {\int_{-\ii}^{+\ii}}
\newcommand {\intoi} {\int_{0}^{+\ii}}

\newcommand {\ee} {\varepsilon}
\newcommand {\lm} {\lambda}

\renewcommand {\t} {\tau}
\newcommand {\om} {\omega}

\newcommand {\muo} {\mu_{0}}
\newcommand {\eo} {\ee_{0}}
\newcommand {\er} {\ee_{\mathrm{r}}}
\newcommand {\ecr} {\ee_{\mathrm{cr}}}
\newcommand {\etao} {\eta_{0}}

\newcommand {\ko} {k_{0}}
\newcommand {\kr} {k_{\rho}}
\newcommand {\kzw} {k_{\rho}^{\mathrm{ZW}}}
\newcommand {\kzo} {k_{z0}}
\newcommand {\kzu} {k_{z1}}
\newcommand {\ku} {k_{1}}
\newcommand {\ks} {k_{\sigma}}

\newcommand {\srm} {\mathrm{s}}
\newcommand {\prm} {\mathrm{p}}
\newcommand {\crm} {\mathrm{c}}

\newcommand {\ra} {\rightarrow}

\newcommand{\kzoq}{\sqr{ 2 \jrm \, \ko \, q + q^{2}}}
\newcommand{\kzuq}{\sqr{ \ku^{2} - \pt{ \ko - \jrm q }^{2}}}
\newcommand{\kzuqu}{\sqr{\,2\jrm\ku q + q^{2}}}
\newcommand{\krqu}{ \pt{ \ku - \jrm q } }

\begin{document}
	
	\title{Pulsed Vertical Electric Dipole Over a Lossy Halfspace: On the Time-Domain Zenneck Wave}

\author{Giampiero Lovat
	\thanks{G. Lovat is with the Electrical Engineering Division of DIEE, University of Rome ``Sapienza'', Rome, via Eudossiana, 18 - 00184 Italy.}
}

\maketitle

\begin{abstract}
	We investigate the transient electromagnetic field radiated by a pulsed vertical electric dipole above a lossy half-space and identify its time-domain signatures associated with the Zenneck wave. Starting from the classical Sommerfeld representation, we derive a causal time-domain formulation based on the double-deformation technique, with successive contour deformations in the transverse-wavenumber and frequency planes. This yields an explicit decomposition of the field into source-pole, loss-pole, modal-pole, and residual steepest-descent contributions.
	The resulting expressions exactly satisfy causality and are validated against a reference solution obtained through a standard double inverse transform. The analysis shows that one modal contribution, generated by the frequency-plane deformation and related to the frequency-domain Zenneck pole, exhibits reduced-time invariance and a spatial attenuation consistent with a surface-wave component. Under suitable source and observation conditions, this term can dominate the field over a broad and physically relevant finite late-time interval.
	At the same time, for the considered damped-sinusoidal excitation, the strict asymptotic tail at fixed distance remains algebraic of order \(t^{-5/2}\), with contributions from both the residual continuous spectrum and the modal-pole family. These results provide a rigorous and physically interpretable time-domain manifestation of the frequency-domain Zenneck wave in the pulsed half-space problem.
\end{abstract}


\section{Introduction}

\IEEEPARstart{S}urface waves excited at planar interfaces play a key role in a wide range of applications, from classical radio propagation to modern plasmonics. Among these, the Zenneck wave (ZW) remains a subject of conceptual and practical interest, having sparked a century-long debate about its physical significance and mathematical interpretation.

The concept of surface-wave (SW) propagation along a planar interface dates back more than a century, to the fundamental works of Sommerfeld and Zenneck. In 1907, Zenneck~\cite{zenneck1907} described a TM-polarized SW solution that could propagate along the interface between air and a lossy medium, showing exponential decay in the vertical direction. This result was extended by Sommerfeld~\cite{sommerfeld1909}, who rigorously solved the problem of a vertical electric dipole (VED) radiating above a conductive half-space. The solution, given in the spectral domain as an integral over the radial wavenumber, exhibited a rich structure that included contributions from continuous and discrete spectral components and, among them, a pole corresponding to the ZW. Despite the elegance of the solution, its interpretation was soon challenged. In 1919, Weyl~\cite{weyl1919} provided an alternative representation for the field of a point source over the ground, using a Green's function formulation that emphasized continuous-spectrum contributions and excluded the Zenneck pole. This discrepancy highlighted a subtle but essential point: the presence of a pole in the spectral integrand does not guarantee that the relevant residue contributes to the physical field, unless the contour encloses it. Thus a longstanding debate began on whether the ZW is a physical phenomenon or a mathematical artifact.

This issue was further examined in the 1930s by Norton~\cite{norton1937}, who extended the asymptotic analysis of the Sommerfeld integral and clarified that, for a realistic Earth ground, the dominant contribution to the field at large distances is not the ZW, but a lateral wave arising from the branch cut of the spectral integrand, known also as \emph{Norton wave}. This component exhibits an algebraic decay and defines the classical groundwave in long-range radio propagation. Norton's correction of Sommerfeld's asymptotics (later recast in a more elegant form by Fock using saddle-point methods~\cite{fock1945}) marginalized the Zenneck contribution for decades. A detailed historical and technical reexamination of these early controversies was later reported by Collin~\cite{collin2004controversies}.

In a series of works~\cite{wait1957}, Wait advanced the idea that the ZW, though mathematically present in the spectral representation, does not play a physically significant role in typical groundwave problems. This led to describe the ZW as \emph{nonphysical}, with particular reference to configurations involving elementary dipole sources.  Related skeptical interpretations have also reappeared more recently in \cite{sarkar2014schelkunoff,sarkar2017clarification}, where the terminology surrounding ZW and SWs was revisited with particular care.

However, not all researchers agreed with this verdict. A number of studies \cite{hill1978zenneck} considered specific source configurations that could selectively excite the ZW. While these efforts did not fully overturn the prevailing idea, they planted the seeds for a later reevaluation. Meanwhile, a growing number of researchers started to re-examine more deeply the Sommerfeld solution taking care of poles, branch points, and Riemann surfaces in the spectral plane. One of the most important contributions in this renewed investigation was the series of papers by Michalski and Mosig~\cite{michalski2015redux,michalski2016redux}. Revisiting the Sommerfeld half-space problem with full analytical rigor, they emphasized that the ZW is not merely a formal residue, but a legitimate solution (corresponding to a simple pole in the complex spectral plane) whose contribution to the field must be assessed through a carefully constructed integration path. Their analyses clarified the topology of the associated Riemann surfaces, the location of the Zenneck pole relative to branch cuts, and the need for proper analytic continuation when interpreting modal contributions. Michalski and Mosig established also a connection between the location of the Zenneck pole and the Brewster angle \cite{michalski2022brewster} which allowed for recasting the ZW as a form of inverted leaky wave~\cite{jackson2022leaky}. In parallel, Jackson and Mesa developed technically robust papers exploring the nature of the ZW in canonical and engineered systems~\cite{mesa2020excitation,jackson2022leaky}. Their contributions showed that, although the ZW may not be readily excited by a point dipole, it can become dominant when the source is appropriately matched. 

These modern interpretations thus showed that the observable relevance of the Zenneck contribution depends on the chosen field representation, on whether the pole is actually captured by the contour deformation, and on the source \emph{and} observation point, as also observed in \cite{bhattacharyya2019longitudinal}  where a longitudinal spectral formulation was used.

Despite this progress, the ZW remains primarily studied as a frequency-domain (FD) component. Transient analyses of this simple configuration are rare \cite{haddad1981transient,kooij1996electromagnetic} and the question of how to isolate the Zenneck contribution from a pulsed excitation remains unanswered in the literature and motivates the present work.

\subsection{Scope and Contribution of This Work}

When one turns to the time domain (TD), the problem becomes substantially more delicate. In fact the TD field is obtained through an inverse Fourier transform and an isolated pole contribution cannot in general be carried over directly without violating causality \cite{tsang1979modified,haddad1981transient}. The central issue is therefore not whether a Zenneck-related pole exists in the spectral representation, but whether its imprint can be extracted from the transient response in a way that is both causal and analytically well defined. This raises the following question: how can one isolate, within the total field radiated by a pulsed dipole, a TD contribution that can be rigorously associated with the Zenneck-wave physics?

An interesting TD viewpoint is offered by the double-deformation technique (DDT), originally introduced by Tsang and Kong to treat transient sources in layered media through a pair of contour deformations in the $\kr$- and  $\om$–planes \cite{tsang1979modified}: in particular, the method yields causal expressions that separate modal residues from the continuous spectrum  \cite{tsang1979modified, ezzeddine1982time,poh1986transient}. Recently, the DDT has been applied to the excitation of TD surface plasmon polaritons on graphene \cite{burghignoli2018time}.

In this paper, we apply the DDT to the transient field radiated by a \emph{pulsed} VED over a lossy half-space and derive a fully causal TD representation of the Sommerfeld solution. The proposed formulation separates the total field into source-pole, loss-pole, modal-pole, and residual steepest-descent contributions, thereby making it possible to identify a Zenneck-related TD modal contribution without resorting to a naive inverse transform of the classical FD pole term. The resulting expressions are validated against an accurate double inverse transform reference. The main point is that numerical results show that one dominant modal contribution exhibits reduced-time invariance and a spatial attenuation consistent with a SW component and with the FD ZW attenuation constant, and that, under suitable source and observation conditions, this contribution can govern a broad and physically relevant finite late-time interval. At the same time, for the considered damped-sinusoidal excitation, the analysis shows that the strict asymptotic tail for $t \to \infty$ at a fixed observation point is algebraic of order $t^{-5/2}$, with contributions from both the residual continuous spectrum and the modal-pole terms. The remainder of this paper is organized as follows. Section II formulates the pulsed VED-over-half-space problem and derives the spectral representation after the deformation in the transverse-wavenumber plane. Sections III--V analyze the subsequent deformation in the complex-frequency plane in three causal time regions and identify the various field contributions. Section VI presents the numerical validation and discusses the TD signatures and possible late-time dominance of the Zenneck-related modal contribution. Finally, Sec. VII summarizes the main conclusions.

\section{VED Over a Lossy Halfspace}\label{sec:ved}

We thus aim at investigating the transient electromagnetic field generated by a VED placed at a planar boundary separating two homogeneous and isotropic media. The VED is represented through the FD density current
\begin{equation}
\mathbf J(\mathbf r,\omega) = \uv{z} I(\omega)\delta(x) \delta(y) \delta(z) ,
\end{equation}
where $I(\omega)=\mathcal{I}(\omega) \ell$ represents the Fourier transform of the TD electric moment $i\pt{t} \ell$.

\begin{figure}[t]
	\centering
	\includegraphics[width=\columnwidth]{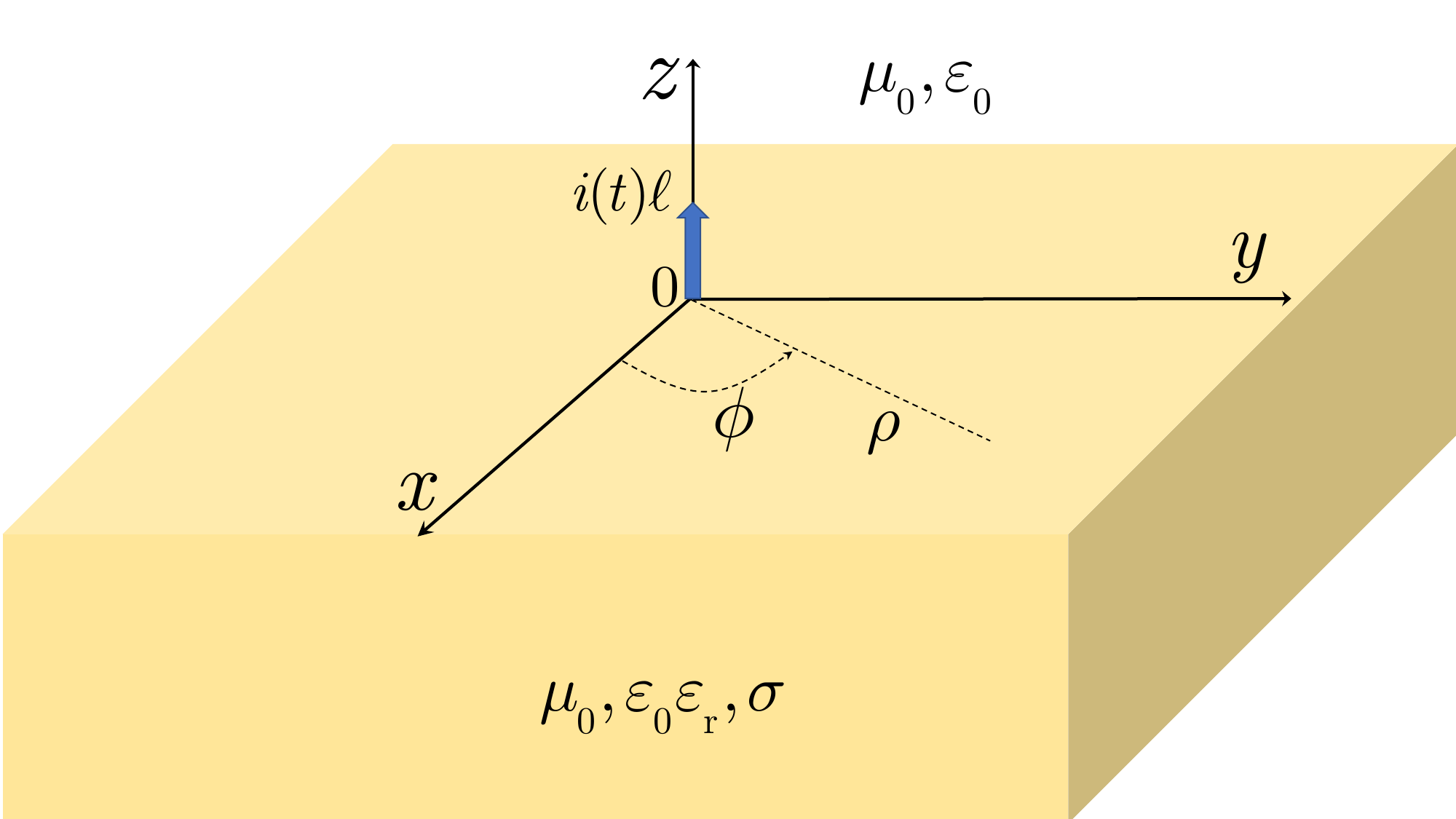}
	\caption{VED at the interface between two half-spaces: geometrical configuration of the problem.}
	\label{fig:VED_geom}
\end{figure}

The configuration is depicted in Fig. \ref{fig:VED_geom}. The upper half-space ($z > 0$) is free space, characterized by permittivity $\eo$ and permeability $\muo$. The lower half-space ($z < 0$) is a lossy medium with complex permittivity $\ee_1 = \ecr \eo$ and the same permeability $\muo$. The relative permittivity is modeled as
\begin{equation} \label{eqn:crec}
	\ecr \pt{\om} = \er - \jrm \fr{\ks }{k_0} ,
\end{equation}
where $\er > 1$, $\ks=\sigma \etao$ (with $\sigma > 0$ and $\etao = \sqr{\muo/\eo}$), and $k_i = \omega \sqr{\muo \ee_i}$ ($i=0,1$).

The radiated field is purely TM and the magnetic field is purely azimuthal. Assuming an $\esp{\jrm \omega t}$ time behavior, its FD spectrum can be reduced to the classical Sommerfeld integral and for $z=0$ we have
\begin{equation}\label{eqn:hf}
	\begin{split}
	&H_\phi(\rho;\omega) = -\jrm \fr{ I\pt{\om}   \ecr\pt{\om}}{4\pi}\\
	&\cdot \intii \fr{  \kr^2 }{ \pq{\ecr\pt{\om} \kzo\pt{\kr, \om}+\kzu\pt{\kr, \om}}}  \bh_1^{(2)}(\kr \rho)  \dd \kr ,
	\end{split}
\end{equation}
where $\bh_1^{(2)}\pt{\cdot}$ denotes the first-order Hankel function of the second kind. The spectral integral in \eqref{eqn:hf} is evaluated along the extended Sommerfeld integration path (ESIP) in the complex $ \kr $-plane as in Fig. \ref{fig:SIP_deformation}. The transverse and vertical wavenumbers are denoted by $\kr$ and $k_{zi}$ ($i=0,1$), respectively, where
\begin{equation} \label{eqn:2.2.3}
	\begin{split}
		k_{zi}\pt{\kr, \om} = \pm \sqrt{k_i^2 - \kr^2} = \pm \sqrt{\om^2 \muo \ee_i - \kr^2} .
	\end{split}
\end{equation}

In the TD, the magnetic field may be expressed as
\begin{equation} \label{eqn:2.2.2b}
	h_\phi\pt{\rho;t} = \re{ \fr{1}{\pi} \int_{0}^{\ii} H_\phi\pt{\rho; \om} \esp{\jrm \om t} \dd \om } .
\end{equation}
The evaluation of the resulting double integral is difficult since the integrand in \eqref{eqn:hf} converges only in the Abel sense.

\subsection{Deformation in the $\kr$ plane}

The Sommerfeld integral in \eqref{eqn:hf} involves the vertical wavenumbers
$k_{zi}(\kr,\omega)$ in \eqref{eqn:2.2.3}, which introduce square-root branch points at $\kr=\pm k_0$ and $\kr=\pm k_1$. Accordingly, the pair $(k_{z0},k_{z1})$ is defined on a \emph{four-sheeted} Riemann surface. Each sheet is labeled by $(S_0,S_1)$, where $S_i=+1$ if $\im{k_{zi}}<0$ and $S_i=-1$ if $\im{k_{zi}}>0$ \cite{MesaJackson2022}. The physical (proper) sheet is $(1,1)$, which makes the field decay away from the interface in both media: the original ESIP lies entirely on $(1,1)$ and avoids the branch cuts \cite{MesaJackson2022}.

\begin{figure}[t]
	\centering
	\includegraphics[width=\columnwidth]{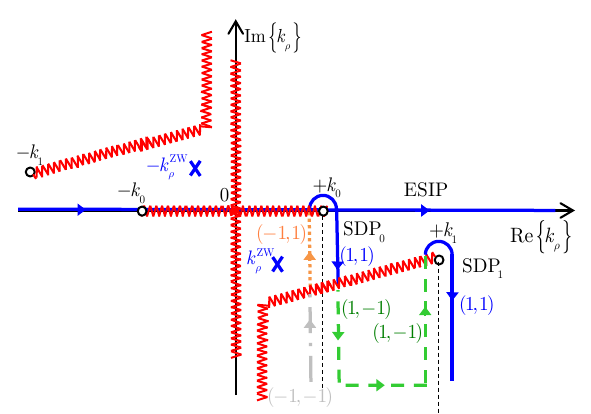}
	\caption{Original and deformed integration paths in $\kr$ plane. The Zenneck pole lies on the $(1,1)$ sheet.}
	\label{fig:SIP_deformation}
\end{figure}

To obtain a rapidly convergent representation, the ESIP is deformed into two steepest-descent paths (SDPs), denoted as $\mathrm{SDP}_0$ and $\mathrm{SDP}_1$, emerging from the branch points at $\kr=k_0$ and $\kr=k_1$, respectively (each consisting of two vertical legs into the lower half-plane), as shown in Fig. \ref{fig:SIP_deformation}. During this deformation the contour crosses Sommerfeld branch cuts and therefore enters different sheets; this is accounted for by evaluating the integrand with the appropriate determinations of $k_{z0}$ and $k_{z1}$ along each segment.

The integrand in \eqref{eqn:hf} also has a simple pole at $\kr=\kzw$, defined by the dispersion relation
\begin{equation}\label{eqn:spole}
	\ecr\,k_{z0}(\kr,\omega)+k_{z1}(\kr,\omega)=0
\end{equation}
from which the FD Zenneck pole is obtained as
\begin{equation}\label{eqn:kzw}
	\kzw=\beta_{\rho}^{\mathrm{ZW}}-\jrm \alpha_{\rho}^{\mathrm{ZW}}=\ko \sqr{\fr{\ecr}{\ecr+1}} .
\end{equation}
For $\re{\ecr}>0$ this yields a pair of solutions $\pm\kzw$. Because of the square roots, the same pole satisfies \eqref{eqn:spole} on more than one sheet (in particular, it appears on $(1,1)$ and on $(-1,-1)$). Moreover, $\kzw$ lies to the left of the line $\re{\kr}=k_0$ \cite{michalski2016redux}, so it is not enclosed by the SDP deformation and does not explicitly enter the standard SDP representation, although it may influence the integrand locally when close to the branch point at $k_0$ \cite{michalski2016redux,MesaJackson2022}.

After the deformation, the field is written as
\begin{equation}\label{eqn:g1}
	H_\phi(\rho;\omega)=H_{\phi 0}(\rho;\omega)+H_{\phi 1}(\rho;\omega)
\end{equation}
with
\begin{equation}\label{eqn:hsdp}
	H_{\phi i}(\rho;\omega)=\int_{\mathrm{SDP}_i}\td{H}_{\!\phi}(\kr,\omega)\,\dd\kr,\qquad i=0,1
\end{equation}
and
\begin{equation}\label{eqn:g2zm}
	\td{H}_{\!\phi}(\kr,\omega)=
	-\jrm\,\frac{I(\omega)}{4\pi}\,
	\frac{\ecr(\omega)\,\kr^2\,\bh_1^{(2)}(\kr\rho)}
	{\ecr(\omega)\,k_{z0}(\kr,\omega)+k_{z1}(\kr,\omega)} .
\end{equation}

Using the parametrization $\kr=k_0-\jrm q$ ($q\ge 0$), the two legs of $\mathrm{SDP}_0$ sample different sheets depending on whether the line $\kr=k_0-\jrm q$ has crossed the $k_1$-branch cut. It can easily be shown that the crossing occurs where $\im{k_{z1}^2}=0$, i.e., for $q=q_0=\ks/2$.
Therefore the $\mathrm{SDP}_0$ contribution can be expressed as the difference of the integrand evaluated on the corresponding sheet pairs, i.e.,
\begin{equation}\label{eqn:g3}
	\begin{split}
		\int_{\mathrm{SDP}_0} \!\!\! \!\!\td{H}_{\!\phi}(\kr,\omega)\dd\kr \!
		&=\! \jrm \!\int_{0}^{q_0}\!\left[\!
		\left.\td{H}_{\!\phi}\right|_{\kr=k_0-\jrm q}^{(-1,1)} \! - \! \!
		\left.\td{H}_{\!\phi}\right|_{\kr=k_0-\jrm q}^{(1,1)}
		\right]\!\dd q\\
		&\!\!\!\!\!\!\!\!+\jrm \int_{q_0}^{\infty}\!\left[
		\left.\td{H}_{\!\phi}\right|_{\kr=k_0-\jrm q}^{(-1,-1)}-
		\left.\td{H}_{\!\phi}\right|_{\kr=k_0-\jrm q}^{(1,-1)}
		\right]\dd q .
	\end{split}
\end{equation}
Using the sheet-dependent substitutions for $k_{z0}$ and $k_{z1}$ we obtain
\begin{equation}\label{eqn:g3a}
	\int_{\mathrm{SDP}_0}\td{H}_{\!\phi}(\kr,\omega)\,\dd\kr
	=\int_{0}^{\infty}\td{H}_{\!\phi 0}(q,\omega)\,\dd q,
\end{equation}
where
\begin{equation} \label{eqn:ht0h}
	\begin{split}
		&	\td{H}_{\! \phi 0}\pt{q,\om}  =-\fr{I\pt{\om} }{2 \pi}\\
		&\cdot
		\dfr{ \ecr^2 \kzoq \pt{k_0-\jrm q}^{2} \bh_1^{\pt{2}}\!\pq{ \pt{k_0 - \jrm q} \rho }}{
			\pt{\ecr-1}\,
			\pq{ \ko^{2}- 2\jrm\ko \pt{\ecr+1} q - \pt{\ecr+1}q^{2} }}  .
	\end{split}
\end{equation}

We next consider the path deformation around the $k_1$ branch point. The contribution along $\mathrm{SDP}_1$ is expressed as the difference between the integrand evaluated on sheet $\pt{1,1}$ and that on sheet $\pt{1,-1}$ with the common parametrization $\kr = \ku - \jrm q$ so that
\begin{equation} \label{eqn:g7a}
	\begin{split}
		\int_{\mathrm{SDP}_1} \td{H}_{\! \phi}\pt{ \kr,\om}  \dd \kr= \int_{0}^{\ii} \td{H}_{\! \phi 1} \pt{q,\om} \dd q ,
	\end{split}
\end{equation}
where
\begin{equation} \label{eqn:g7b}
	\begin{split}
		\td{H}_{\! \phi 1}\pt{q,\om}  =
		\jrm  \pg{\left. \td{H}_{\! \phi} \right|_{\kr=\ku-\jrm q}^{(1,-1)}- \left. \td{H}_{\! \phi} \right|_{\kr=\ku-\jrm q}^{(1,1)} } 
	\end{split}
\end{equation}
which can be similarly shown to be
\begin{equation} \label{eqn:ht1h}
	\begin{split}
		&\td{H}_{\! \phi 1}\pt{q,\om}=\fr{I\pt{\om} }{2\pi} \\
		&\cdot \fr{\ecr \kzuqu \krqu^2 \bh_1^{\pt{2}}\pq{\krqu \rho}}{\pt{\ecr-1} \pq{\ecr^2\ko^2 -2 \jrm \ko  \sqr{\ecr} \pt{\ecr+1}q - \pt{\ecr+1}q^2}} .
	\end{split}
\end{equation}

After deforming the integration path in the $\kr$-plane, the total azimuthal magnetic field is therefore
\begin{equation}\label{eqn:hphi}
	h_\phi(\rho,t)=h_{\phi 0}(\rho,t)+h_{\phi 1}(\rho,t) .
\end{equation}
By the change of variable $k_0=\omega/c$ (with $c=1/\sqr{\mu_0\varepsilon_0}$), letting $\tau = ct$ and $\hat{I}(k_0)=c I\pt{c \ko } $, and using explicitly \eqref{eqn:crec}, we have
\begin{equation}\label{eqn:2.2.19}
	\begin{split}
		&h_{\phi 0}\pt{\rho,\tau}= \mathrm{Re} \left\{\!- \fr{1}{2\pi^2} \! \!    \int_{0}^{\infty} \hat{I}\pt{\ko} \fr{\pt{\er \ko - \jrm \ks }^2}{\pq{\pt{\er-1} \ko - \jrm \ks }} \esp{\jrm \ko \tau} \right.\\
		&\left.\int_0^{\infty}  \fr{\pt{\ko-\jrm q}^2 \sqr{2\jrm q \ko +q^2} \bh_1^{(2)}\left[\pt{\ko-\jrm q}\rho\right]}{D_0\pt{\ko,q}}   \dd q \dd \ko \right\} ,
	\end{split}
\end{equation}
where
\begin{equation}\label{eqn:2.2.22}
	D_0\!\pt{\ko,q}\!\!=\!\ko^3\!-\!2\jrm q \pt{\er\!+ \!1}\ko^2\!-\!q\pq{2 \ks\!\! + q\pt{\er\! +1}}\!\ko \!+\!\jrm q^2 \ks 
\end{equation}
and
\begin{equation}\label{eqn:2.2.20}
	\begin{split}
		&h_{\phi 1}\pt{\rho,\tau} = \mathrm{Re} \left\{ \fr{1}{2\pi^2}\!\!   \int_{0}^{\infty} \hat{I}\pt{\ko} \fr{\er \ko - \jrm \ks }{\pq{\pt{\er-1} \ko - \jrm \ks }} \esp{\jrm \ko \tau} \right.\\
		&\left. \int_0^{\infty}  \fr{\pt{\ku-\jrm q}^2 \sqr{2 \jrm q \ku+q^2}\bh_1^{(2)}\left[\pt{\ku-\jrm q}\rho\right]}{D_1\pt{\ko,q}}   \dd q \dd \ko \right\} ,
	\end{split}
\end{equation}
where
\begin{equation}\label{eqn:2.2.31}
	D_1\pt{\ko,q}=\ku^2\ecr-2\jrm q \ku \pt{\ecr+1}-q^2 \pt{\ecr+1} .
\end{equation}

\subsection{Deformation in the $\ko$ plane}

Now we deform the integration path in the complex $ k_0 $-plane, shifting the contour along the positive real axis onto an SDP which, for the considered configuration, coincides with the imaginary axis. The deformation must respect the radiation condition, which determines the use of the positive or negative part of the imaginary axis. The correct choice is made according on how the integrand behaves for large values of $\abs{\ko}$. Using the large-argument expression of $\bh_1^{(2)}\pt{x}$,
 the asymptotic forms of the integrands in \eqref{eqn:2.2.19} and \eqref{eqn:2.2.20} are
 $\esp{-\jrm \ko \pt{\rho-\tau}}
 $ 
 and 
 $
 \esp{-\jrm \ko \pt{\rho \sqr{\er}-\tau}}
 $,
 respectively. Therefore, the choice of the positive or negative imaginary axis is dictated by the ordering of $\tau$ and $\rho$. In particular, for $\tau<\rho$ both contributions $h_{\phi 0}$ and $h_{\phi 1}$ are rotated to the negative imaginary axis. For $\rho<\tau<\rho\sqrt{\er}$, the contour for $h_{\phi 0}$ must be taken to the positive imaginary axis, whereas the contour for $h_{\phi 1}$ remains on the negative imaginary axis. Finally, for $\tau>\rho\sqrt{\er}$, both contours are rotated to the positive imaginary axis.

 The original real‐axis contour is thus replaced by an arc in the first or fourth quadrant, plus a vertical leg along the imaginary axis. Jordan's lemma ensures that the arc contribution vanishes, so that we remain with the double integrals over $q$ and $\ko$ on the imaginary axis \emph{and} the residues of any poles enclosed by the deformation. These poles may originate from the poles of $\hat{I}\pt{k_0}$ (\emph{source poles}), from the zero of the term $\pq{\pt{\er-1} \ko - \jrm \ks }$ (\emph{loss pole}) or from the zeros of the polynomial  \eqref{eqn:2.2.22} and \eqref{eqn:2.2.31} (\emph{modal poles}).  
\begin{figure}[t]
	\centering
	\includegraphics[width=\columnwidth]{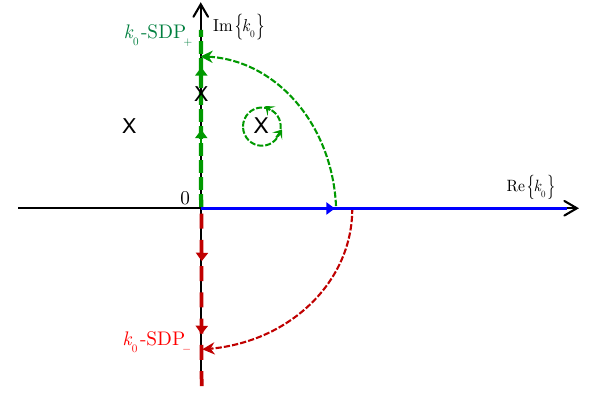}
	\caption{Original integration path (blue solid line) and $\ko$-SDPs for $k_0$ integration. Dashed red line: $\ko$-SDP for SDP$_0$ and SDP$_1$ (when $\tau<\rho$), $\ko$-SDP for SDP$_1$ (when $\tau < \rho \sqr{\er}$). Dashed green line: $\ko$-SDP for SDP$_0$ (when $\tau > \rho$) and SDP$_1$ (when $\tau > \rho \sqr{\er}$).}
	\label{fig:k0_paths}
\end{figure}

\section{Time Behavior for $ \rho > \tau$}\label{sec:t1}

For $\tau<\rho$, the $\ko$-integration path is deformed to $\text{SDP}_{-}$ ($k_0=-\jrm u$, $u>0$), as illustrated in Fig.~\ref{fig:k0_paths}, thus obtaining
	\begin{equation}\label{eqn:2.2.6}
		\begin{split}
			&h_{\phi 0 s}\pt{\rho,\tau}= \mathrm{Re} \left\{ \jrm \fr{1}{2\pi^2}     \int_{0}^{\infty} \hat{I}\pt{-\jrm u} \fr{\ecr^2}{\pt{\ecr-1}} \esp{ u \tau} \right.\\
			&\left.\int_0^{\infty}  \fr{ \pt{u+ q}^2 \sqr{\pt{u+ q}^2- u^2} \bh_1^{(2)}\left[-\jrm \pt{ u+ q}\rho\right]}{ u^2 +2 u q \pt{\ecr+1} +q^2 \pt{\ecr+1}}   \dd q \dd u \right\} ,
		\end{split}
	\end{equation}
where $\ecr=\ecr\pt{-\jrm u}$. Since a physically realizable current source produces a spectrum $\hat{I}\pt{\ko}$ that is analytic in the fourth quadrant, no \emph{source poles} are encountered.

From $\ecr(k_0)-1=0$, a single \emph{loss pole} can be identified on the \emph{positive} imaginary axis located at
\begin{equation}\label{eqn:lp}
	\ko^{\sigma}= \jrm\fr{\ks}{\er-1}.
\end{equation}	

Finally, since all the zeros of the denominators reside in the upper half of the $k_{0}$-plane for every $q$ and $\ko$ (as can rigorously be proven by letting $\ko=-\jrm \lm$ and verifying that the resulting polynomial with real coefficients satisfy the Hurwitz criterion \cite{Bodson2020}), the path deformation encounters no \emph{modal poles}.

Now, $\ecr\pt{-\jrm u}$ is real and strictly positive and, for a physically realizable current source, $\hat{I}\pt{-\jrm u}$ is real as well. In addition, since
$
	\bh_1^{(2)}\pt{-\jrm x}=-\fr{2}{\pi}\bk_{1}\pt{x},
$
for $x>0$ (
where $\bk_{1}\pt{\cdot}$ is the first-order modified Bessel function of the second kind), the factor $\bh_1^{(2)}\pq{-\jrm \pt{u+q}\rho} $ 
is real for $u>0$ and $q>0$.  Consequently, from \eqref{eqn:2.2.6} it results $h_{\phi 0 s}=0$.
	
Analogously, we have
	\begin{equation}\label{eqn:hf1s}
		\begin{split}
			&h_{\phi 1 s}\pt{\rho,\tau} = \mathrm{Re} \left\{- \jrm\fr{1}{2\pi^2}   \int_{0}^{\infty} \hat{I}\pt{-\jrm u} \fr{\ecr}{\pt{\ecr-1}} \esp{u \tau}  \right.\\
			&\quad  \int_0^{\infty}  \fr{\pt{u \sqr{\ecr}+ q}^2 \sqr{\pq{ u \sqr{\ecr}+ q}^2-u^2 \ecr} }{u^2\ecr^2+ 2  q u\sqr{\ecr}\pt{\ecr+1}+q^2 \pt{\ecr+1}}    \\
				&\quad \left.\fan{\int_0^{\infty}}  \cdot \bh_1^{(2)}\left[-\jrm \pt{ u \sqr{\ecr}+ q} \rho\right]   \dd q \dd u \right\}
		\end{split}
	\end{equation}	
and with a similar reasoning it can be shown that $h_{\phi 1 s}=0$.

Hence the total field vanishes identically for all $t<\rho/c$, in full agreement with causality.

\section{Time Behavior for $ \rho < \tau < \rho \sqr{\er} $}\label{sec:ht2}

For $\rho<\tau<\rho\sqrt{\er}$, convergence of the $k_0$-integrals is obtained by rotating the contour associated with $\mathrm{SDP}_0$ onto the positive imaginary axis ($k_0=\jrm u$), while the contour associated with $\mathrm{SDP}_1$ is rotated onto SDP$_{-}$.

In this time interval, the argument illustrated in the previous section implies that the $\mathrm{SDP}_1$ term does not contribute, so that $h_{\phi 1}\equiv 0$. The total field can thus be decomposed as
\begin{equation}\label{eqn:htot0}
	h_{\phi 0}(\rho,\tau)=
	h_{\phi 0}^{\srm}(\rho,\tau)+
	h_{\phi 0}^{\sigma}(\rho,\tau)+
	h_{\phi 0}^{\prm}(\rho,\tau)+
	h_{\phi 0}^{\mathrm{SDP}}(\rho,\tau).
\end{equation}
In \eqref{eqn:htot0}, $h_{\phi 0}^{\srm}$ is the contribution of the source poles, $h_{\phi 0}^{\sigma}$ that of the loss pole $k_0^\sigma$, $h_{\phi 0}^{\prm}$ is the total contribution of the modal poles captured during the deformation, and $h_{\phi 0}^{\mathrm{SDP}}$ is the remaining SDP double integral over $q$ and $u$.

\subsection{Source-pole contribution $h_{\phi 0}^{\srm}$}
\label{sec:4.1}

If $\hat I(\ko)$ has an $(n\!+\!1)$th--order pole at $\ko=\ko^\srm$ in the upper half $\ko$--plane, the source--pole contribution $h_{\phi0}^{\srm}$ is
\begin{equation}\label{eq:hs0}
	\begin{split}
		&h_{\phi0}^{\srm} \pt{\rho,\t}
		= \mathrm{Re}\left\{- \fr{\jrm \epsilon}{\pi n!}
			\lim_{\ko\ra \ko^\srm}
			\fr{\dd^n}{\dd \ko^n}
			\left[
			\pt{\ko-\ko^\srm}^{n+1}\,\esp{\jrm \ko \t}
			 \right. \right.
			\\
			&\qquad \cdot \fr{\hat I(\ko) \pt{\ko\er-\jrm \ks }^2}{\ko(\er-1)-\jrm \ks } \\
			&\left. \left.  \int_{0}^{\ii}
			\fr{\pt{\ko-\jrm q}^{2}\sqr{\jrm q\pt{2\ko-\jrm q}}}
			{D_0(\ko,q)}
			\bh_1^{(2)}\pq{\pt{\ko-\jrm q}\rho}\dd q
			\right] 
		\right\} .
	\end{split}
\end{equation}
It should be pointed out that the $ \ko^\srm $ pole
contributes to the field only if $\re{\ko^\srm} \geq 0$.  Therefore, $\epsilon=1$ if the pole $\ko^\srm$ has a positive real part, $\epsilon=1/2$ if the pole $\ko^\srm$ is purely imaginary, and $\epsilon=0$ if the pole $\ko^\srm$ has a negative real part.

\subsection{Loss-pole contribution $h_{\phi 0}^{\sigma}$}

The loss pole in \eqref{eqn:lp}
lies on the deformed path on the positive imaginary axis and is independent of $q$, so that its contribution to $ h_{\phi 0} $ is
\begin{equation}\label{eqn:lp1}
	\begin{split}
		&h_{\phi 0}^{\sigma}\pt{\rho,\tau} = \mathrm{Re} \left\{ -\jrm \fr{ \hat{I} \pt{\ko^{\sigma}}  \esp{\jrm \ko^{\sigma} \tau}}{2 \pi \pt{\er-1}}      \pt{\ko^{\sigma} \er -\jrm \ks }^2 \right. \\
		&  \left.  \int_0^{\ii} \fr{\pt{\ko^{\sigma}-\jrm q}^2 \sqr{\jrm q \pt{2 \ko^{\sigma} -\jrm q}}}{D_0\pt{\ko^{\sigma},q}} \bh_{1}^{(2)} \pq{\pt{\ko^{\sigma}-\jrm q} \rho} 	\dd q\right\} . 
	\end{split}
\end{equation}
Since
	\begin{equation}\label{eqn:d0s}
	D_0\pt{\ko^{\sigma},q}=-\jrm \fr{\sigma^3 \etao^3}{\pt{ \er - 1}^3}+4 \jrm q  \fr{\sigma^2 \etao^2}{ \pt{\er - 1}^2}  -2\jrm q^2 \fr{\ks }{ \er - 1} ,
\end{equation}
using $ \gamma = \ks  / (\er - 1) $ we have
	\begin{equation}\label{eqn:d1s}
	D_0\pt{\ko^{\sigma},q}=-2 \jrm \gamma\pt{q^2-2 q  \gamma  +\fr{\gamma^2}{2}} ,
\end{equation}
i.e., two poles are present  at $
	q_{1,2}=\gamma \pt{1\pm 1/\sqr{2}}
$.
Using the change of variable $ \xi = q - \gamma $ 
we have
\begin{equation}\label{eqn:hpa1}
	\begin{split}
	&h_{\phi 0}^{\sigma}\pt{\rho,\tau} = \mathrm{Re} \left\{  \fr{1}{4 \pi} \esp{-\gamma \tau} \hat{I}\pt{\jrm \gamma}  \fr{\gamma}{\er-1} \right. \\
	&\left. \int_{-\gamma}^{\ii}     \dfr{\xi^2 \sqr{\xi^2-\gamma^2}}{\pt{\xi-\gamma/\sqr{2}}\pt{\xi+\gamma/\sqr{2}}} \bh_{1}^{(2)}\pt{-\jrm \xi \rho}        \dd \xi  \right\} 
	\end{split}
\end{equation}
and the poles are located at
$\xi = \pm \gamma/\sqr{2}$, i.e., inside the range of integration. The integration interval can be separated as $[-\gamma,\gamma]$ and $[\gamma, +\ii)$ and it can be noted that for $\xi > \gamma$ no poles are present and the integral is purely real since the square-root is real, $\hat{I}\pt{\jrm \gamma}$ is real, and the Hankel function is real as well.
Therefore
\begin{equation}\label{eqn:hpa3}
	\begin{split}
		&h_{\phi 0}^{\sigma}\pt{\rho,\tau}
		= \re{  \fr{\jrm}{4 \pi} \esp{-\gamma \tau} \hat{I}\pt{\jrm \gamma}  \fr{\gamma}{\er-1} \int_{-\gamma}^{\gamma}    F\pt{\xi}        \dd \xi  } \\
			&-  \fr{1}{2 \pi^2} \esp{-\gamma \tau} \hat{I}\pt{\jrm \gamma}  \fr{\gamma}{\er-1} \int_{\gamma}^{\ii}    \xi^2 \dfr{ \sqr{\xi^2-\gamma^2}}{\xi^2-\gamma^2/2}  \bk_{1}\pt{\xi \rho}        \dd \xi   ,
	\end{split}
\end{equation}
where
\begin{equation}\label{eqn:fbeta}
	\begin{split}
 F\pt{\xi}	=    \dfr{ \xi^2 \sqr{\gamma^2-\xi^2}}{\pt{\xi-\gamma/\sqr{2}}\pt{\xi+\gamma/\sqr{2}}} \bh_{1}^{(2)}\pt{-\jrm \xi \rho}     .   
	\end{split}
\end{equation}

Because of the real poles  at $\xi = \pm \gamma/\sqr{2}$, the effects of such singularities are included through their residues. In particular,
\begin{equation}
	\begin{split}
	\mathrm{Res} \pq{F\pt{\xi}}_{\xi=\pm \gamma/\sqr{2}}= \pm \fr{\gamma^2}{4}\bh_{1}^{(2)}\pt{\mp\jrm \fr{\gamma}{  \sqr{2}} \rho} 
	\end{split}
\end{equation}
and after regularizing  the integral we finally obtain
\begin{equation}\label{eqn:finl0a}
	\begin{split}
		&	h_{\phi 0}^{\sigma}\pt{\rho,\tau}=	\mathrm{Re} \left\{ \fr{\jrm}{4 \pi} \esp{-\gamma \tau} \hat{I}\pt{\jrm \gamma}  \fr{\gamma}{\er-1} \right. \\
		& \cdot \int_{-\gamma}^{\gamma}\left[    F\pt{\xi} -\fr{\gamma^2}{4}\bh_{1}^{(2)}\pt{-\jrm \fr{\gamma}{  \sqr{2}} \rho} \fr{1}{\xi - \gamma / \sqr{2}} \right. \\  
		&\fan{aaaaa}\left. \left. +\fr{\gamma^2}{4}\bh_{1}^{(2)}\pt{\jrm \fr{\gamma}{  \sqr{2}} \rho} \fr{1}{\xi + \gamma / \sqr{2}}  \right]  \dd \xi \right\} \\
		& -\fr{\esp{-\gamma \tau}}{8 \pi}  \hat{I}\pt{\jrm \gamma}  \fr{\gamma^3}{\er-1} \ln \pt{\fr{\sqr{2}+1}{\sqr{2}-1}}    \bi_{1}\pt{\fr{\gamma}{  \sqr{2}} \rho}     \\
		&- \fr{\esp{-\gamma \tau}}{2 \pi^2}  \hat{I}\pt{\jrm \gamma}  \fr{\gamma}{\er-1} \int_{\gamma}^{\ii}    \dfr{\xi^2  \sqr{\xi^2-\gamma^2}}{\xi^2-\gamma^2/2}  \bk_{1}\pt{\xi \rho}        \dd \xi   ,
	\end{split}
\end{equation}
having used the identity $	\bh_{1}^{(2)}\pt{\jrm x}  +  \bh_{1}^{(2)}\pt{-\jrm x} = 2\jrm \bi_{1}\pt{x}$, where $\bi_{1}\pt{\cdot}$ is the first-order modified Bessel function of the first kind.

\subsection{Modal-pole contributions $h_{\phi 0}^{\prm}$}\label{sec:opc0}

To determine the contribution associated with the zeros in $k_0$ of $D_0(k_0,q)$, we have to track the root trajectories as functions of the real parameter $q>0$ and check whether they are intercepted by the $k_0$-contour deformation. The three roots are obtained by solving $D_0(k_0,q)=0$. For any fixed $q>0$, the set $\{k_{0}^{(1,m)}\}_{m=1}^3$ falls into one of the following configurations:
\begin{enumerate}
	\item all the roots are distinct on the positive imaginary axis;
	\item  all the roots are purely imaginary and positive, with two coincident (this degeneracy can occur only at isolated values of $q$ and therefore it does not affect the subsequent integral representations);
	\item  one root is purely imaginary and positive, while the other two have equal imaginary parts and opposite real parts.
\end{enumerate}
Finally, note that for $q=0$ the denominator $D_0(k_0,0)$ has a third-order zero at $k_0=0$ where the integrand vanishes.

When we have a set of three distinct roots on the positive imaginary axis, the contribution to the field can be written as
\begin{equation}
	h_{\phi 0}^{\mathrm{p}}\pt{\rho,\tau}=\sum_{m=1}^{3} h_{\phi 0 }^{\prm m}\pt{\rho,\tau} ,
\end{equation}
where
\begin{equation}\label{eqn:2.2.25}
	\begin{split}
	&	h_{\phi 0}^{\prm m}\pt{\rho,\tau}= \mathrm{Re} \left\{ - \frac{\jrm }{2 \pi} \intoi \esp{\jrm k_0^{(1,m)} \tau} \hat{I}\pt{k_0^{(1,m)}} \right. \\
	& \frac{\pt{k_0^{(1,m)}\er-\jrm \ks }^2 \sqrt{\jrm q \pt{2k_0^{(1,m)}-\jrm q}}}{ \pq{k_0^{(1,m)}(\er-1)-\jrm \ks } (k_0^{(1,m)}-k_0^{(1,j)}) (k_0^{(1,m)}-k_0^{(1,n)})} \\
		&\left. \fan{\intoi \esp{\jrm k_0^{(1,m)} \tau}} \cdot (k_0^{(1,m)}-\jrm q)^2   \bh_{1}^{(2)}[(k_0^{(1,m)}-\jrm q) \rho] \dd q \right\} ,
	\end{split}
\end{equation}
where $ j $ and $ n $ are either $ 1$, $2$ or $3 $ with $ j\neq m\neq n $.

When there is a simple pole $k_0^{\srm \prm}$ on the positive imaginary axis and two symmetric poles on the first ($k_0^{\crm 1}$) and second ($k_0^{\crm 2}$) quadrants of the $k_0$ plane, only $k_0^{\crm 1}$ and $k_0^{\srm \prm}$ are intercepted in the deformation. We thus have 
\begin{equation}
	h_{\phi 0}^{\mathrm{p}}\pt{\rho,\tau}= h_{\phi 0 }^{\mathrm{sp}}\pt{\rho,\tau}+h_{\phi 0}^{\mathrm{cp}}\pt{\rho,\tau} ,
\end{equation}
where
\begin{equation}\label{eqn:2.2.28}
	\begin{split}
		&h_{\phi 0}^{\mathrm{sp}}(\rho,\tau)= \mathrm{Re} \left\{ - \frac{\jrm }{2 \pi} \intoi \esp{\jrm k_0^{\srm \prm} \tau} \hat{I}(k_0^{\srm \prm}) (k_0^{\srm \prm}-\jrm q)^2 \right. \\
	 &\left.\frac{ (k_0^{\srm \prm}\er-\jrm \ks )^2 \sqrt{\jrm q (2k_0^{\srm \prm}-\jrm q)}\; \bh_{1}^{(2)} [(k_0^{\srm \prm}-\jrm q) \rho]}{\pq{k_0^{\srm \prm}(\er-1)-\jrm \ks }(k_0^{\srm \prm}-k_0^{\crm 1})(k_0^{\srm \prm}-k_0^{\crm 2})}  \dd q \right\} 
	\end{split}
\end{equation}
and
\begin{equation}\label{eqn:2.2.29}
	\begin{split}
	&	h_{\phi 0}^{\mathrm{cp}}(\rho,\tau)= \mathrm{Re} \left\{ - \frac{\jrm }{ \pi} \intoi \esp{\jrm k_0^{\crm 1} \tau} \hat{I}(k_0^{\crm 1}) (k_0^{\crm 1}-\jrm q)^2 \right. \\
	 &\left. \frac{(k_0^{\crm 1}\er-\jrm \ks )^2  \sqrt{\jrm q \pt{2k_0^{\crm 1}-\jrm q}} \; \bh_{1}^{(2)} [(k_0^{\crm 1}-\jrm q) \rho] }{\pq{k_0^{\crm 1}(\er-1)-\jrm \ks }(k_0^{\crm 1}-k_0^{\crm 2}) (k_0^{\crm 1}-k_0^{\srm \prm})}    \dd q \right\} .
	\end{split}
\end{equation}

\subsection{Double-integral contribution $h_{\phi 0}^{\mathrm{SDP}}$}

The double integral contribution after the deformation of SDP$_0$ to the positive imaginary $k_0$ axis ($k_0 = \jrm u$) is
\begin{equation}\label{eqn:2.2.30}
	\begin{split}
	&	h_{\phi 0}^{\mathrm{SDP}}\pt{\rho,\tau} = \mathrm{Re} \left\{ -\fr{\jrm}{2 \pi^2} \! \! \intoi \! \! \dd q \; \mathrm{PV} \!\! \int_{0}^{\ii} \! \! \fr{\esp{-u \tau} \hat{I}\pt{\jrm u}}{\pq{u\pt{\er-1}-\ks} } \right. \\
		&\left. 
		 \fr{\pt{u \er-\ks }^2 \pt{q-u}^2 \sqr{q\pt{q-2u}} \bh_{1}^{(2)}\pq{\jrm \pt{u-q} \rho}}{u^3-2q \pt{\er+1}u^2+q\pq{2 \ks + q\pt{\er+1}}u-q^2\ks }  \dd u 	\right\} .
		\end{split}
\end{equation}
The $u$ integration is a Cauchy principal-value integral avoiding all the poles possibly located along the integration path. By considering the integrand in \eqref{eqn:2.2.30}, it should be noted that for $u$ and $q$ real $\hat{I}\pt{\jrm u}$ is real, while $\bh_{1}^{(2)}\pq{\jrm \pt{u-q} \rho}$ is real for $u<q$; on the other hand, $\sqr{q\pt{q-2u}}$ is real if $u<q/2$ and purely imaginary for $u>q/2$. Therefore, since for $u<q/2$ the integrand is purely imaginary, we have
\begin{equation}\label{eqn:2.2.30c}
	\begin{split}
		&h_{\phi 0}^{\mathrm{SDP}}\pt{\rho,\tau} = \mathrm{Re} \left\{ \fr{1}{2 \pi^2} \! \! \intoi \! \! \dd q \; \mathrm{PV} \int_{q/2}^{\ii} \fr{\esp{-u \tau} \hat{I}\pt{\jrm u} }{\pq{u\pt{\er-1}-\ks }} \right. \\
		&\left.\fr{\pt{u \er-\ks }^2 \pt{u-q}^2 \sqr{q\pt{2u-q}} \bh_{1}^{(2)}\pq{\jrm \pt{u-q} \rho}}{u^3-2q \pt{\er+1}u^2+q\pq{2 \ks + q\pt{\er+1}}u-q^2\ks }  \dd u 	\right\} ,
	\end{split}
\end{equation}
where the lowest limit in the $u$ integration has been set to $q/2$.

Numerically, the PV reconstruction is explicit. First, all real poles on the integration path are identified and the corresponding residues \(R_i\) are computed. Then the regular part is integrated excluding small neighborhoods of the poles and the analytical PV logarithmic term is finally added.

\section{Time Behavior for $\tau > \rho \sqr{\er}$}

For $\tau > \rho \sqr{\er}$, both SDP$_0$ and SDP$_1$ are deformed to the positive imaginary $k_0$ axis ($k_0 = \jrm u$). Consequently, the contributions due to the SDP$_0$ deformation remain the same as those discussed in Sec. \ref{sec:ht2}. In this section we then discuss only the contributions due to the SDP$_1$ deformation. As in the case of $h_{\phi 0}\pt{\rho,\tau}$,  referring  to \eqref{eqn:2.2.20}, $h_{\phi 1}$ may be expressed as
\begin{equation}\label{eqn:htot1}
	h_{\phi 1}\pt{\rho,\tau}=	h_{\phi 1}^{\srm}\pt{\rho,\tau}	+	h_{\phi 1}^{\sigma}\pt{\rho,\tau}+	h_{\phi 1}^{\prm}\pt{\rho,\tau}+	h_{\phi 1}^{\mathrm{SDP}}\pt{\rho,\tau}
\end{equation}
where $h_{\phi 1}^{\srm}$ is the contribution of the source poles of $\hat I(k_0)$, $h_{\phi 1}^{\sigma}$ is the residue associated with the loss pole $k_0^\sigma$, $h_{\phi 1}^{\prm}$ collects the residues of the modal poles generated by the zeros of $D_1(k_0,q)$, and $h_{\phi 1}^{\mathrm{SDP}}$ is the double integral over $q$ and $u$.

\subsection{Source-pole contribution $h_{\phi 1}^{\srm}$}

For $\hat{I}(k_0)$ with an $(n + 1)$-th order pole at $k_0 = \ko^{\srm}$, the source-pole contribution is computed from
\begin{equation}\label{eqn:2.2.32}
	\begin{split}
&\!\!	h_{\phi 1}^{\srm}\pt{\rho,\tau} =  \mathrm{Re} \left\{ -\fr{\jrm \epsilon}{\pi n!} \lim_{\ko \ra \ko^{\srm}} \fr{\dd^n}{\dd \ko^n} \left[   \pt{\ko-\ko^{\srm}}^{n+1}  \right. \right. \\
&\fan{aaaaaaaaaa}\hat{I} \pt{\ko} \esp{\jrm \ko \tau}  \fr{\ko \er -\jrm \ks }{\ko \pt{\er-1}-\jrm \ks }                   \\
	&\!\!\!\!\left. \left. \cdot \int_0^{\ii}\!\! \fr{\pt{\ku-\jrm q}^2 \sqr{\jrm q \pt{2 \ku -\jrm q}} \; \bh_{1}^{(2)} \pq{\pt{\ku-\jrm q} \rho}}{\pq{\ecr \ku^2-2 \jrm q \ku \pt{1+\ecr}-q^2\pt{1+\ecr}}}  \! \dd q\right]\!	\right\} .
\end{split}
\end{equation}

\subsection{Loss-pole contribution $h_{\phi 1}^{\sigma}$}

The loss pole $\ko^{\sigma}$ lies on the positive imaginary $k_0$-axis and it is easy to check that its contribution to $h_{\phi 1}$ may be written exactly as the negative of $h_{\phi 0}^{\sigma}(\tau)$, so that for $\tau > \rho \sqr{\er}$ the total loss-pole contribution is identically \emph{zero}, i.e.,
\begin{equation}\label{eqn:2.2.34}
	h_{\phi}^{\sigma}\pt{\rho,\tau}=h_{\phi 0}^{\sigma}\pt{\rho,\tau}+h_{\phi 1}^{\sigma}\pt{\rho,\tau}=0 .
\end{equation}

\subsection{Modal-pole contributions $h_{\phi 1}^{\prm}$}

From \eqref{eqn:2.2.31} we obtain
\begin{equation}
	\begin{split}
D_1(k_0, q) =& \left(\ko \er - \jrm \ks \right)^2 - q^2 \fr{\pq{\ko \pt{\er+1}-\jrm \ks} }{\ko}\\
&- 2 \jrm q \sqr{\fr{\ko \er-\jrm \ks }{\ko}}\pq{\ko \pt{\er+1}-\jrm \ks } 
\end{split}
\end{equation}
and we then set $D_1(k_0, q) = 0$ to obtain the zeros.

The determination of the contributions of the zeros of $D_1(k_0, q)$  is similar to the description provided in Sec. \ref{sec:opc0} and depends on the location of the zeros in the first quadrant of the $k_0$ plane.

\subsection{Double-integral contribution $h_{\phi 1}^{\mathrm{SDP}}$}

Finally, the double integral contribution  is
\begin{equation}\label{eqn:2.2.35}
	\begin{split}
	&h_{\phi 1}^{\textrm{SDP}}\pt{\rho,\tau} = \mathrm{Re} \left\{  \fr{\jrm}{2 \pi^2} \int_{0}^{\ii} \dd q \; \pv \int_{0}^{\ii} \dd u \esp{-u\tau} \hat{I}\pt{\jrm u}                         \right. \\
&\fr{u \pt{q-\sqr{u^2\er-\ks  u}}^2 \sqr{q \pt{q-2\sqr{u^2 \er-\ks  u}}}}{\hat D_1(u,q)} \\
&\left. \fr{\pt{\er u -\ks }}{\pq{\pt{\er-1}u -\ks }}  \bh_{1}^{(2)} \pq{-\jrm \pt{q-\sqr{u^2\er-\ks  u}} \rho}	\right\} ,
	\end{split}
\end{equation}
where
\begin{equation}
	\begin{split}
		\hat D_1(u,q)=& u \pt{\er u-\ks }^2-2 q \sqr{u^2\er-\ks  u} \pq{\pt{\er+1}u-\ks }\\
		&+q^2 \pq{\pt{\er+1}u-\ks } .
	\end{split}
\end{equation}

The total field for the considered time interval is the sum of contributions from the deformations of both SDP$_0$ and SDP$_1$ and therefore
\begin{equation}\label{eqn:htot100}
	\begin{split}
		h_{\phi}\pt{\rho,\tau}
		=&h_{\phi 0}^{\srm}\pt{\rho,\tau}	+	h_{\phi 0}^{\prm}\pt{\rho,\tau}+	h_{\phi 0}^{\mathrm{SDP}}\pt{\rho,\tau}+h_{\phi 1}^{\srm}\pt{\rho,\tau}\\
		&+	h_{\phi 1}^{\prm}\pt{\rho,\tau}+	h_{\phi 1}^{\mathrm{SDP}}\pt{\rho,\tau} .
	\end{split}
\end{equation}

\section{Numerical Results}
We assume a relative permittivity of the lower half-space $\er = 3.2$ and a conductivity $\sigma = 0.02~\mathrm{S/m}$. 
The dipole current is modeled as a causal pulsed excitation. In particular, we adopt the \emph{damped-sinusoidal} current \cite{tsang1979modified,ezzeddine1982time}
\begin{equation}\label{eqn:scds}
	i(t)=I_0 \fr{t}{T_0} \sin \pt{\omega_0 t} \esp{-\alpha_0 t} H\pt{t} ,
\end{equation}
where $H(t)$ is the Heaviside unit-step function, for which
\begin{equation}\label{eqn:Iw_ds22}
	\hat{I}(k_0)=\frac{2\jrm}{c T_0} I_0 \ell \left[ \frac{\hat{\omega}_0 (k_0-\jrm \hat{\alpha}_0)}{(k_0+\hat{\omega}_0-\jrm \hat{\alpha}_0)^{2} (k_0-\hat{\omega}_0-\jrm \hat{\alpha}_0)^{2}} \right],
\end{equation}
where $\hat{\omega}_0=\omega_0/c$ and $\hat{\alpha}_0=\alpha_0/c$, and $T_0=2 \pi/\omega_0$. In the following examples, $\hat{\omega}_0$ and $\hat{\alpha}_0$ are therefore used as source parameters. It is evident that one double complex source pole exists at
$ k_{0}^{\srm}=\hat{\omega}_0+\jrm \hat{\alpha}_0 $
that contributes to the field through \eqref{eq:hs0} with $n=1$. We choose the central frequency $f_0=\omega_0/(2 \pi)$ below the conduction--displacement transition
$f_{c}=\sigma/(2\pi\eo \er)$ (so $\tan\delta \gtrsim 1$),
while keeping $f$ high enough that the observation point is outside the quasi-static region
(i.e., $\rho > \lambda_0/(2\pi)$). We thus adopt $f_0=30$ MHz and $\hat{\alpha}_0=\hat{\omega}_0/2$.

\subsection{Modal-pole dispersion and DDT contributions}
\label{sec:num_zw}
\begin{figure}[t]
	\centering
	\begin{subfigure}{\columnwidth}
	\includegraphics[width=\columnwidth]{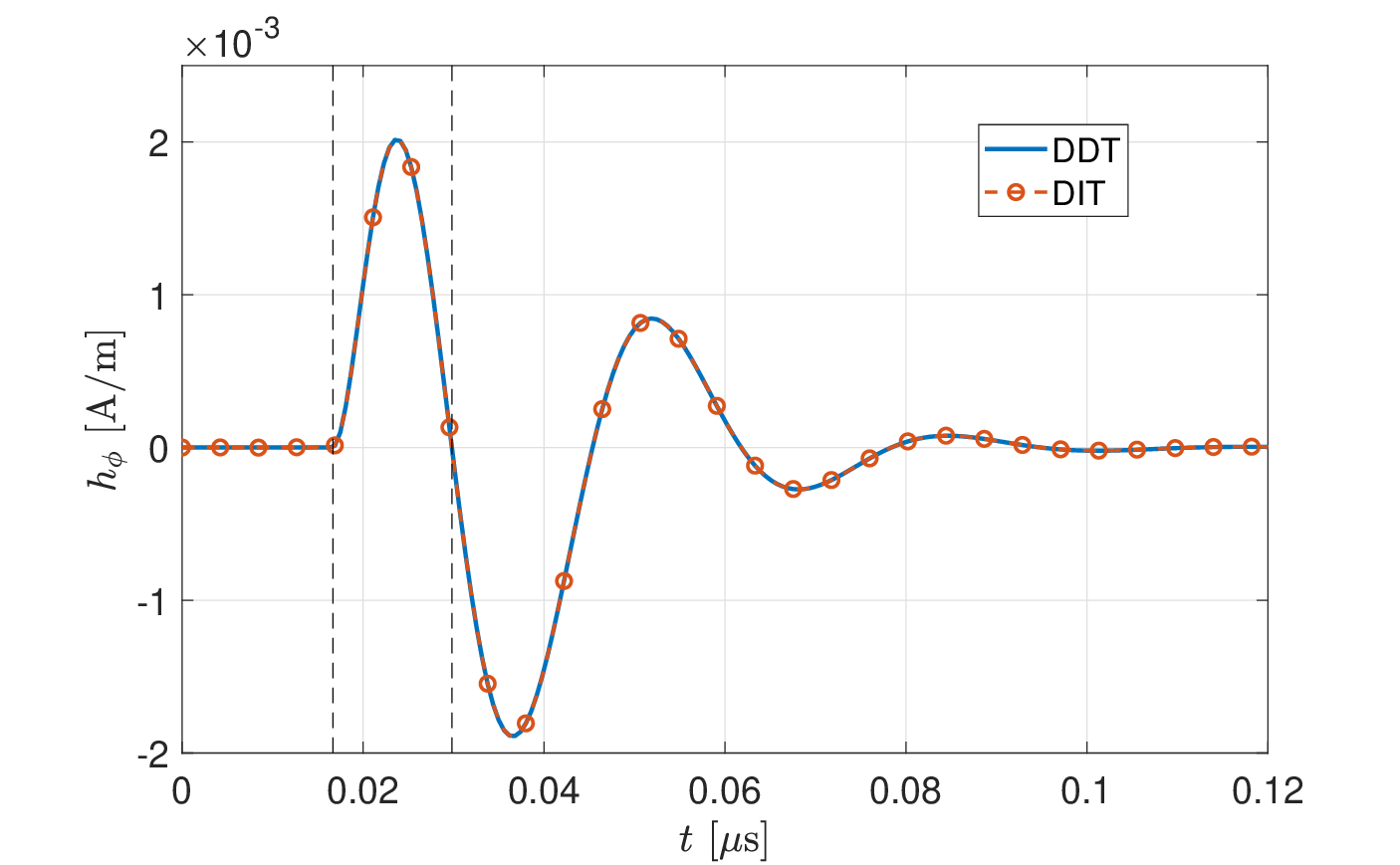}
	\caption{}
	\label{fig:ca}
\end{subfigure}
\begin{subfigure}{\columnwidth}
	\includegraphics[width=\columnwidth]{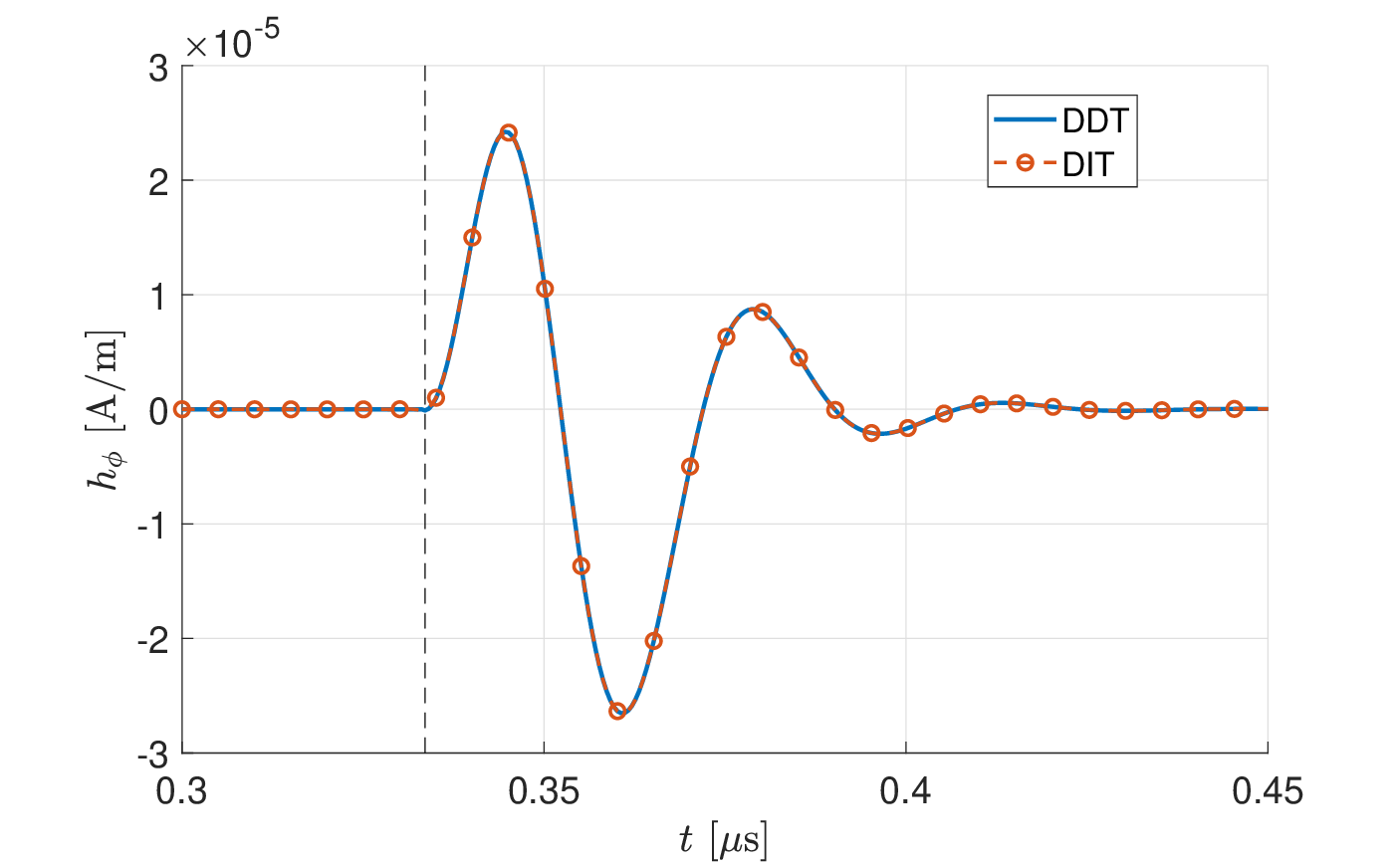}
	\caption{}
	\label{fig:cb}
\end{subfigure}
	\caption{TD field $h_{\phi}$ calculated through the proposed DDT and through a double inverse transform (DIT) for two representative distances: $\rho=5~\mathrm{m}$ (a) and $\rho=100~\mathrm{m}$ (b).}
	\label{fig:decomp_time}
\end{figure}

First of all we check the accuracy of the proposed DDT formulation.
In all TD plots we report the azimuthal magnetic field $h_\phi(\rho,t)$ as a function of the time $t$ at a certain distance $\rho$. We thus first compare the field calculated with the proposed DDT formulation with that obtained through a double inverse transform (DIT) 
accurately calculated with the technique presented in \cite{lovatTEMC2026}. In particular, in Fig.~\ref{fig:decomp_time} the TD field is reported  for two different lateral distances $\rho = 5$ m (intermediate horizontal range) and $\rho=100$ m (far zone).
 As it can be seen, the curves are perfectly superimposed (the lines $t=\rho/c$ and $t=\rho \sqrt{\er}/c$ are also reported). 
 
\begin{figure}[t]
 	\centering
 	 \includegraphics[width=\columnwidth]{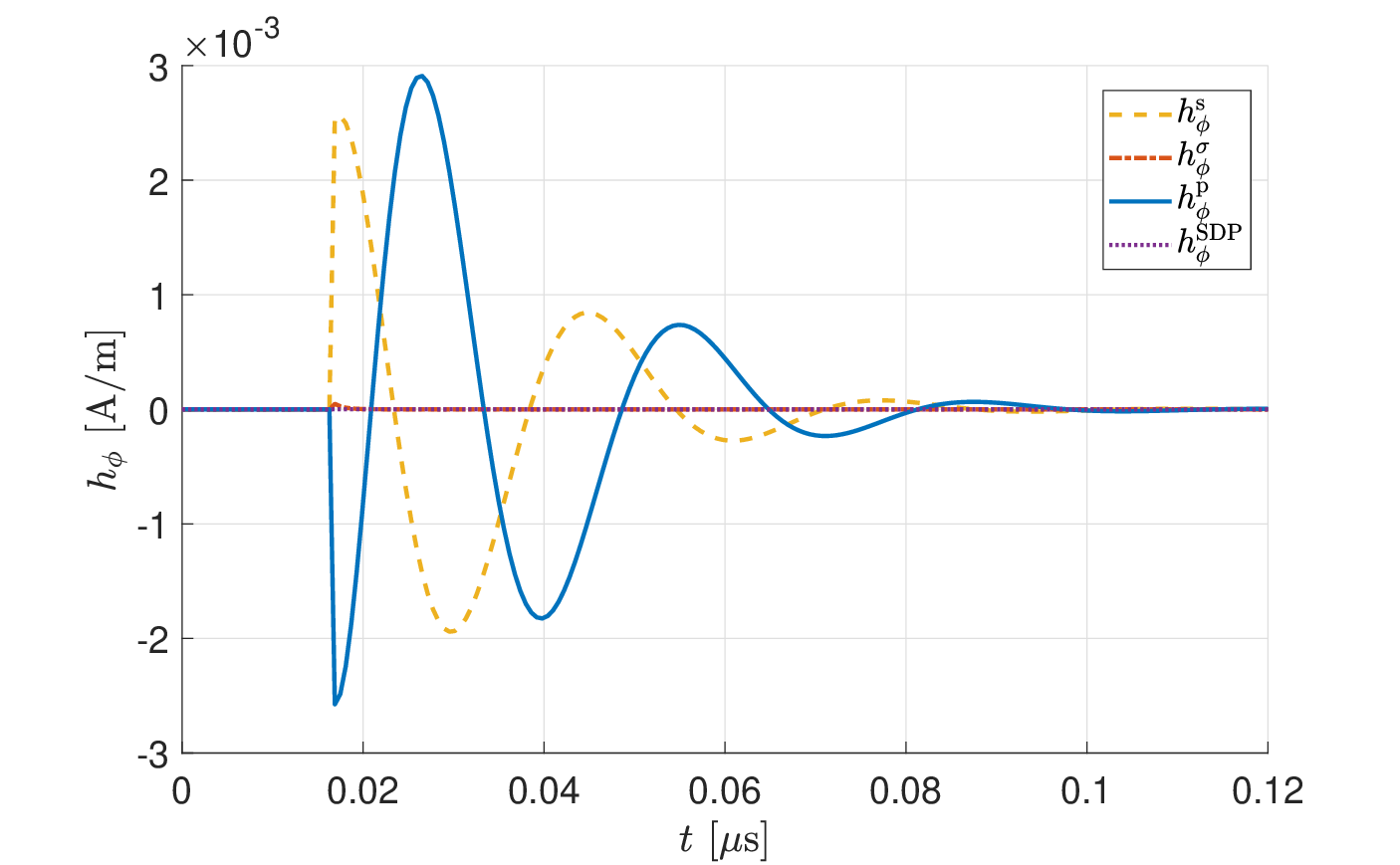}
 	\caption{DDT contributions of $h_\phi(\rho,t)$ for the case $\rho=5~\mathrm{m}$.}
 	\label{fig:ddt_terms}
 \end{figure}

\begin{figure}[t]
	\centering
	\begin{subfigure}{\columnwidth}
		\includegraphics[width=\columnwidth]{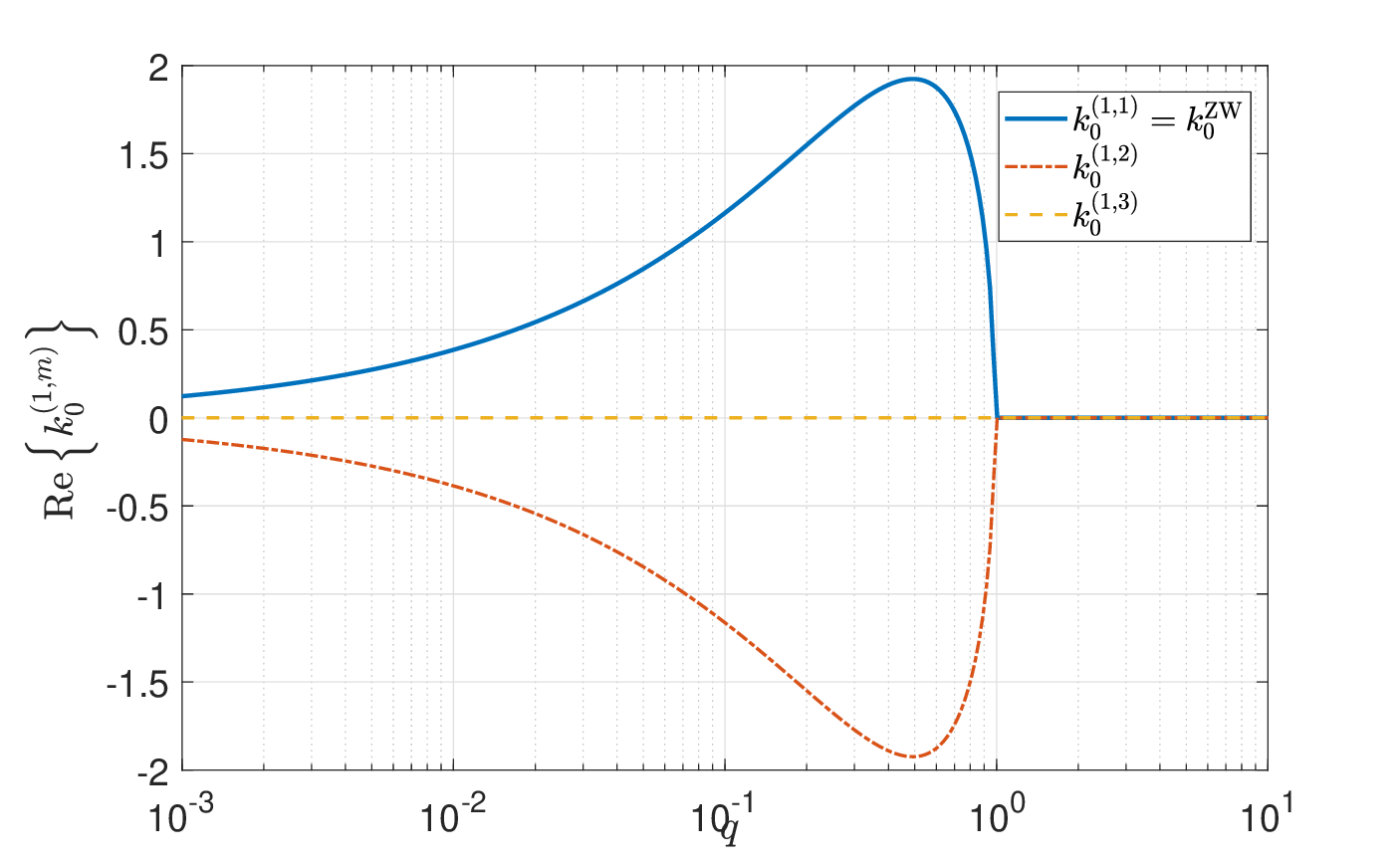}
		\caption{}
		\label{fig:a}
	\end{subfigure}
	\begin{subfigure}{\columnwidth}
		\includegraphics[width=\columnwidth]{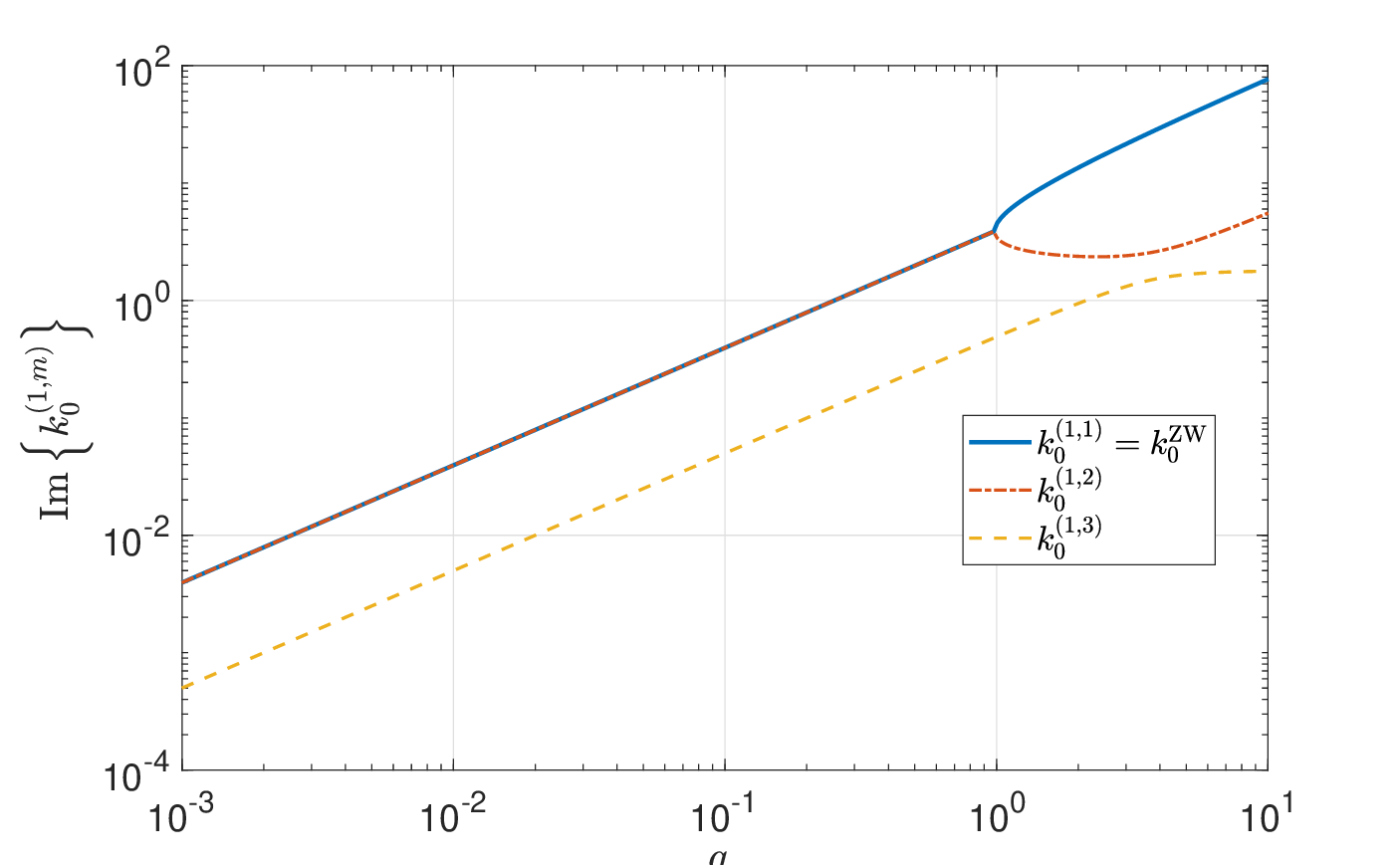}
		\caption{}
		\label{fig:b}
	\end{subfigure}
	\caption{Modal-pole dispersion curves: the three roots $k_0^{(1,m)}(q)$ as functions of the real parameter $q\ge 0$. \emph{Parameters}: $\er=3.2$ and $\sigma=0.02~\mathrm{S/m}$.  Real part (a) and imaginary part (b).}
	\label{fig:dispersion_D0}
\end{figure}

 We are now ready to illustrate the modal content of the DDT formulation and identify the TD signatures associated with the ZW solution.
In the considered case, the dominant contributions are the source-pole term $h_{\phi 0}^{\srm}$ and the modal-pole sum $h_{\phi 0}^{\prm}$, while the remaining terms are much smaller.
 This is clearly evident in Fig. ~\ref{fig:ddt_terms} where we report the various contribution to $h_\phi(\rho,t)$ for the case $\rho=5$ m. 
 Moreover, most notably, among the various modal-pole contributions, \emph{one} modal pole is dominant, the others being completely negligible.
 To this end, in Fig.~\ref{fig:dispersion_D0} we report the dispersion curves $k_0^{(1,m)}(q)$ obtained from $D_0(k_0,q)=0$.
 Although three distinct trajectories exist for each $q$, the modal-pole response  is accurately represented by a \emph{single} dominant modal pole, which we denote as
 $k_0^{\mathrm{ZW}} \equiv k_0^{(1,1)}(q)$.
 This pole is generated by the second (frequency-plane) deformation and, as clarified next, provides a compact and physically interpretable \emph{TD footprint} of the ZW: as it can be seen, $k_0^{\mathrm{ZW}}(q)$ is complex for small $q$ and becomes purely imaginary for larger $q$.
 
\begin{figure}[t]
	\centering
		\begin{subfigure}{\columnwidth}
		\includegraphics[width=\columnwidth]{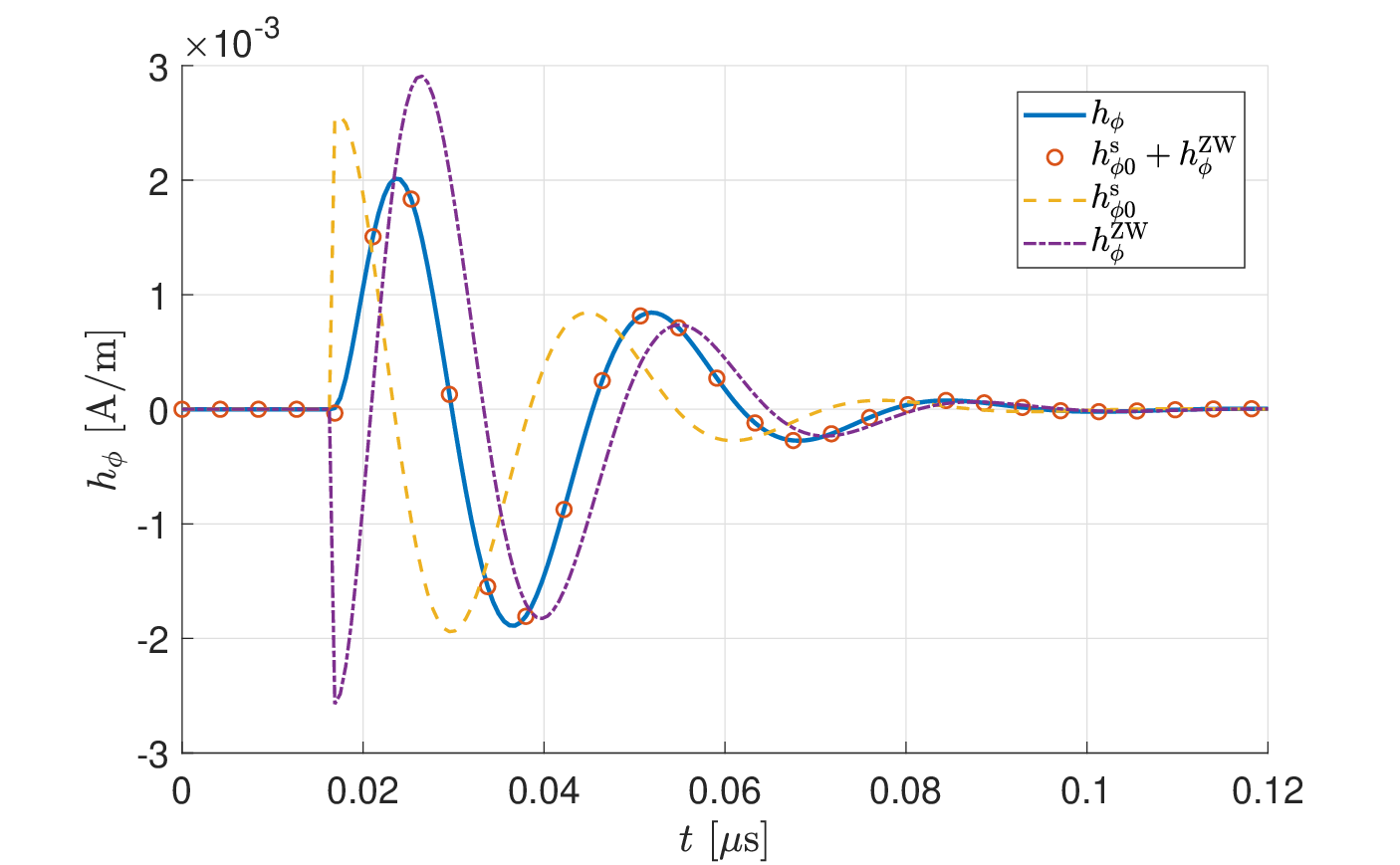}
		\caption{}
		\label{fig:da}
	\end{subfigure}
	\begin{subfigure}{\columnwidth}
		\includegraphics[width=\columnwidth]{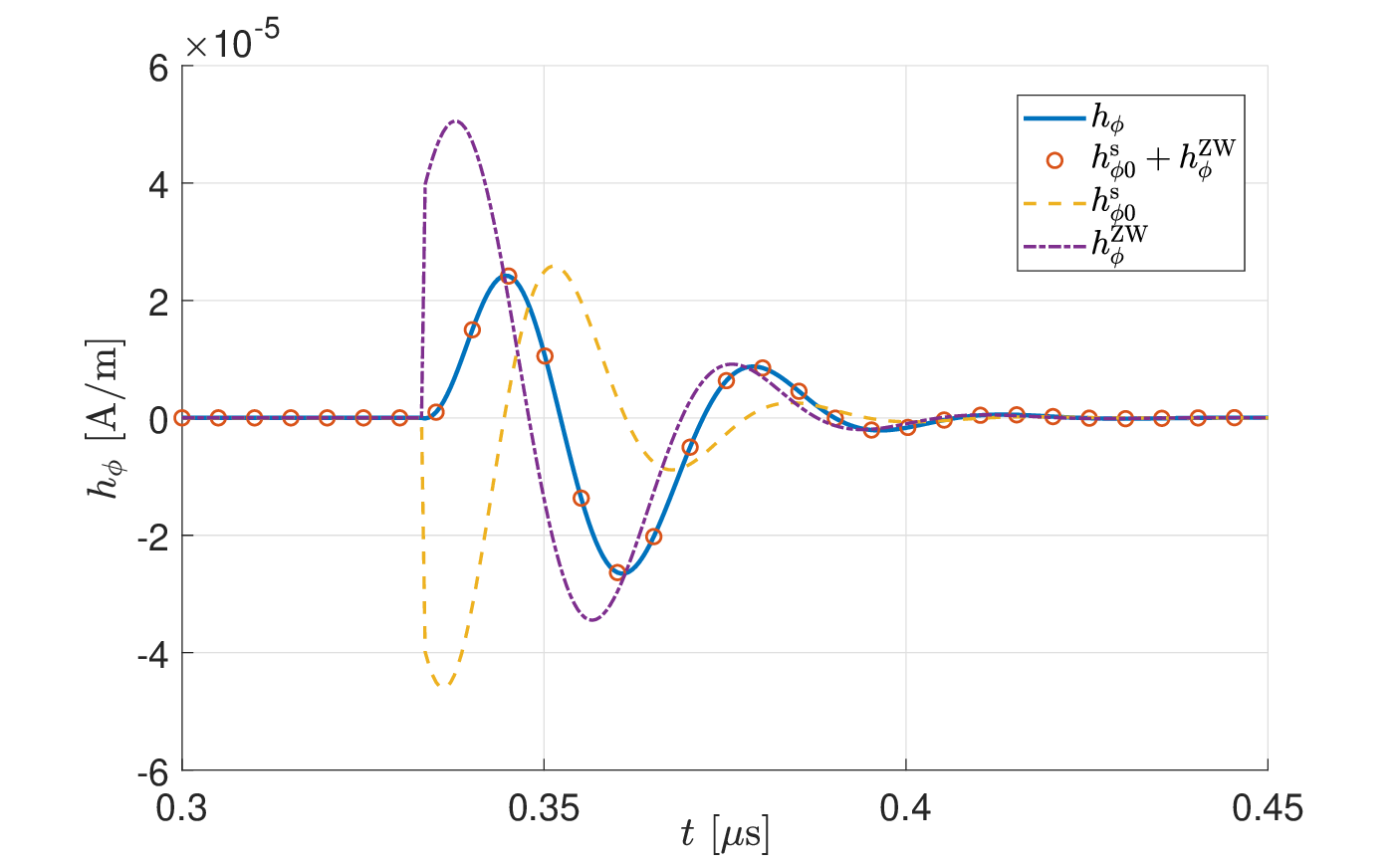}
		\caption{}
		\label{fig:db}
	\end{subfigure}
	\caption{TD field $h_\phi(\rho,t)$ for $\rho=5~\mathrm{m}$ (a) and $\rho=100~\mathrm{m}$ (b): total field $h_{\phi}$ calculated through the proposed DDT, its approximation $h_{\phi 0}^{\srm} + h_{\phi}^{\mathrm{ZW}}$, and the single components $h_{\phi 0}^{\srm}$ and $h_{\phi}^{\mathrm{ZW}}$.}
	\label{fig:decomp_time2}
\end{figure}

In practice, a very accurate approximation of the field is obtained as
\begin{equation}
	h_\phi(\rho,\tau) \simeq  h_{\phi 0}^{\srm}(\rho,\tau) + h_{\phi}^{\mathrm{ZW}}(\rho,\tau),
	\label{eq:reduced_model}
\end{equation}
where $h_{\phi}^{\mathrm{ZW}}=h_{\phi 0}^{\prm 1}$ denotes the single dominant modal contribution due to the pole $k_0^{\mathrm{ZW}}$.
This is clearly illustrated in Fig.~\ref{fig:decomp_time2}, where, for $\rho=5$ m and $\rho=100$ m, we compare the exact field of Fig.~\ref{fig:decomp_time} with the field calculated through \eqref{eq:reduced_model} which is observed to hold with excellent accuracy. For reference, the relevant contributions $h_{\phi 0}^{\srm}$ and $h_{\phi}^{\mathrm{ZW}}$ are also reported: as it can be seen the inclusion of both contributions is essential to recover the total TD waveform.

\subsection{Footprints of the TD Zenneck wave}

We next show that the dominant modal contribution $h_{\phi}^{\mathrm{ZW}}$ exhibits the characteristic space--time behavior of a Zenneck SW component.
It is thus convenient to introduce the reduced-time variable
$ \tau_\rho=\tau-\rho$,
which measures the time elapsed after the arrival of the causal front along the interface.

\begin{figure}[t]
	\centering
	\includegraphics[width=\columnwidth]{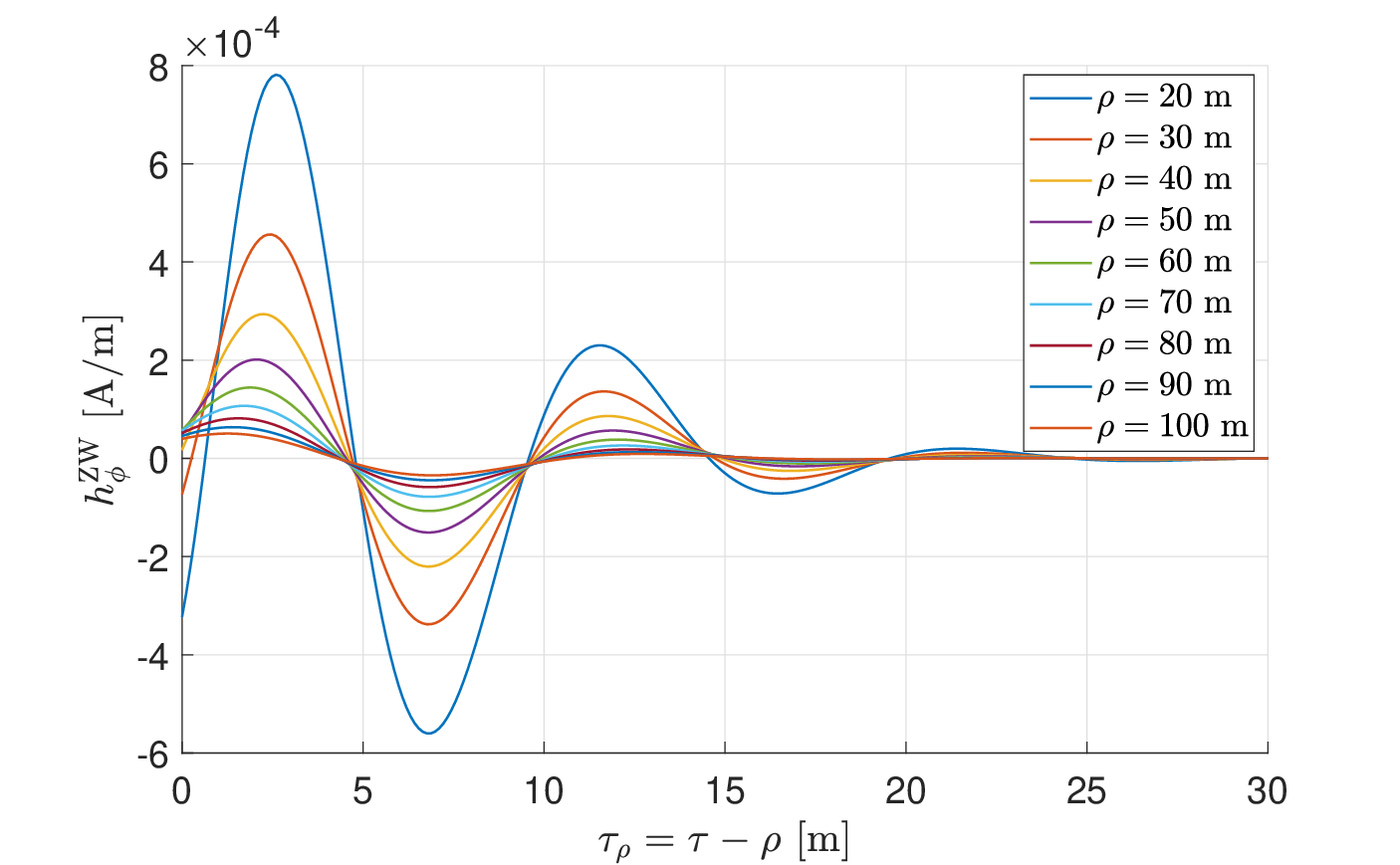}
	\caption{Modal contribution $h_{\phi}^{\mathrm{ZW}}$ as a function of the reduced time $\tau_{\rho}$ for different values of $\rho$.}
	\label{fig:reduced_time_overlay}
\end{figure}

 Figure~\ref{fig:reduced_time_overlay} plots $h_{\phi}^{\mathrm{ZW}}$ as a function of  $\tau_{\rho}$ for several distances.
The waveforms exhibit a near collapse, indicating that $h_{\phi}^{\mathrm{ZW}}$ propagates along the interface with a distance-proportional delay while preserving a largely invariant TD signature.
This behavior provides a first, direct footprint of an interfacial SW component in the TD response.

A second and more quantitative Zenneck-wave (ZW) footprint is provided by the spatial attenuation of the dominant modal contribution $h_{\phi}^{\mathrm{ZW}}$.
For a numerically robust extraction, we focus on the main pulse-like feature of $h_{\phi}^{\mathrm{ZW}}\pt{\tau_\rho,\rho}$ and denote by $\tau_{\mathrm{pk}}(\rho)$ its peak position in reduced time. Introducing the peak-centered variable
$
	\chi=\tau_\rho-\tau_{\mathrm{pk}}(\rho)
$
we examine the waveform within the fixed window $\mathcal W_{\mathrm{pk}}=[-\chi_{0},\chi_{0}]$, with $\chi_0=\lambda_0/2$, chosen with the same width for all $\rho$. At a sufficiently large reference distance $\rho_{\mathrm{ref}}$, where the modal waveform is already asymptotic, the corresponding peak-centered profile is adopted as the reference waveform:
\begin{equation}
	\psi(\chi)=\frac{h_{\phi}^{\mathrm{ZW}}\pq{\rho_{\mathrm{ref}},\tau_{\mathrm{pk}}(\rho_{\mathrm{ref}})+\chi}}
	{\left(\int_{\mathcal W_{\mathrm{pk}}}\left|h_{\phi}^{\mathrm{ZW}}\pq{\rho_{\mathrm{ref}},\tau_{\mathrm{pk}}(\rho_{\mathrm{ref}})+\chi}\right|^2\,\mathrm{d}\chi\right)^{1/2}} .
\end{equation}
The modal amplitude is then defined through the projection coefficient
\begin{equation}
	A(\rho)=\Abs{\int_{\mathcal W_{\mathrm{pk}}}
	h_{\phi}^{\mathrm{ZW}}\pq{\rho,\tau_{\mathrm{pk}}(\rho)+\chi}\,\psi(\chi)\,\mathrm{d}\chi}.
\end{equation}
This definition  provides a stable estimate of the modal strength while preserving the reduced-time signature of the packet.

To test the expected SW attenuation law,
\begin{equation}
	A(\rho)\propto \frac{\exp(-\alpha_\rho \rho)}{\sqrt{\rho}},
\end{equation}
we introduce the quantity $
	y(\rho)=\ln\!\left[A(\rho)\sqrt{\rho}\right]$.
If the modal contribution is in its asymptotic surface-wave regime, $y(\rho)$ should vary approximately linearly with $\rho$, with slope $-\alpha_\rho$.

The attenuation constant is therefore extracted by least-squares linear regression of $y(\rho)$ over an automatically selected asymptotic interval of observation points and the fitted slope directly yields the attenuation constant $\alpha_\rho$, while the corresponding $R^2$ provides a compact measure of how accurately the modal field follows the expected decay law.

\begin{figure}[t]
	\centering
	 \includegraphics[width=\columnwidth]{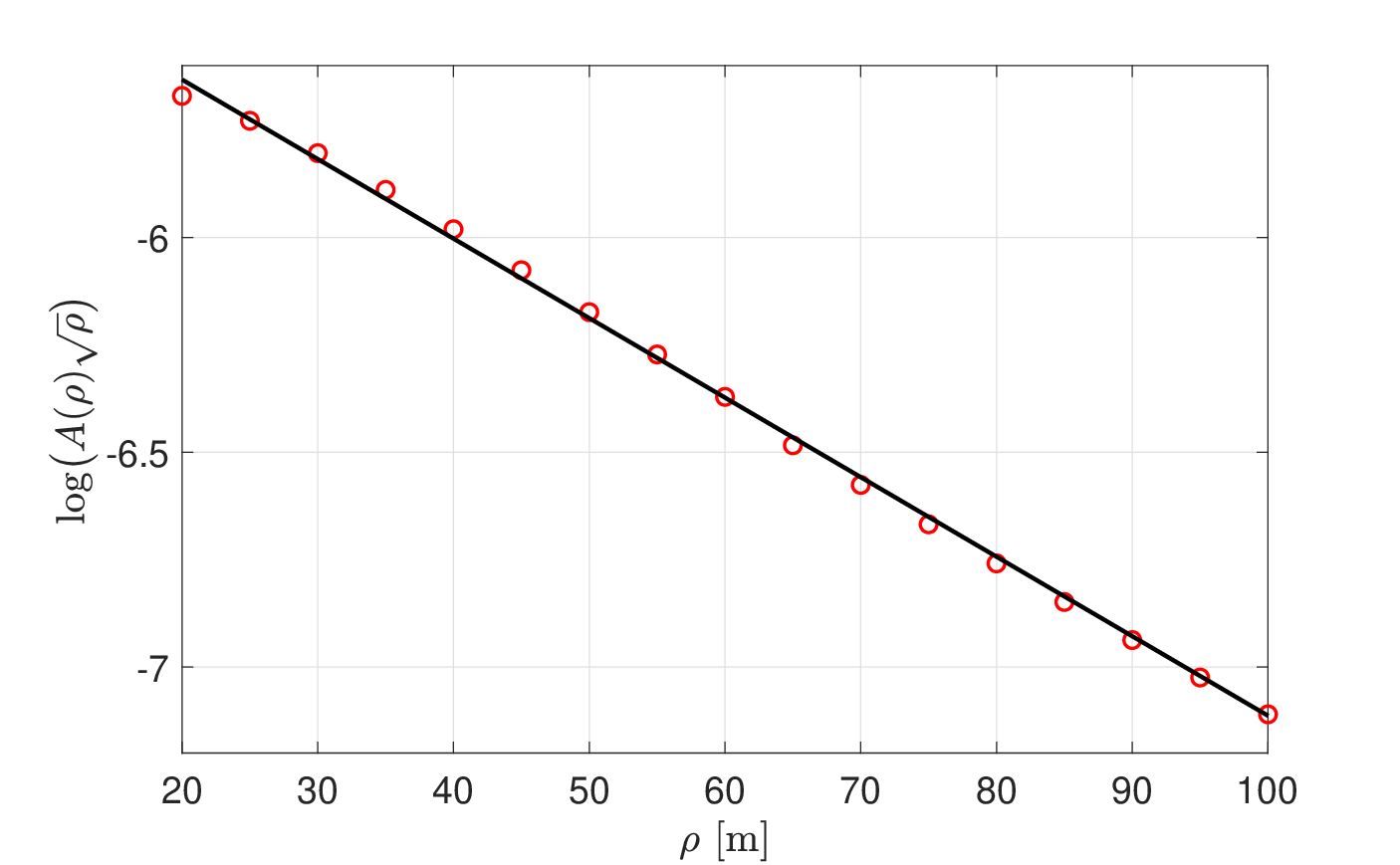}
	\caption{Spatial attenuation of the modal contribution $h_{\phi}^{\mathrm{ZW}}$.}
	\label{fig:attenuation_fit}
\end{figure}

 For the considered structure, the resulting fit, reported in Fig.~\ref{fig:attenuation_fit}, yields an effective attenuation $\alpha_\rho=0.019\,\mathrm{m^{-1}}$ with a coefficient of determination $R^2=0.99$. This attenuation $\alpha_\rho$ is close to the imaginary part of the FD Zenneck pole evaluated at the central frequency, i.e.,  $\alpha_{\rho}^{\mathrm{ZW}}(\omega_0)=0.023\,\mathrm{m^{-1}}$. The observed difference is consistent with the fact that $\alpha_\rho$ is an \emph{effective} attenuation which reflects a weighted contribution of spectral components over the pulse bandwidth.

\begin{figure}[t]
	\centering
	\begin{subfigure}{\columnwidth}
		\includegraphics[width=\columnwidth]{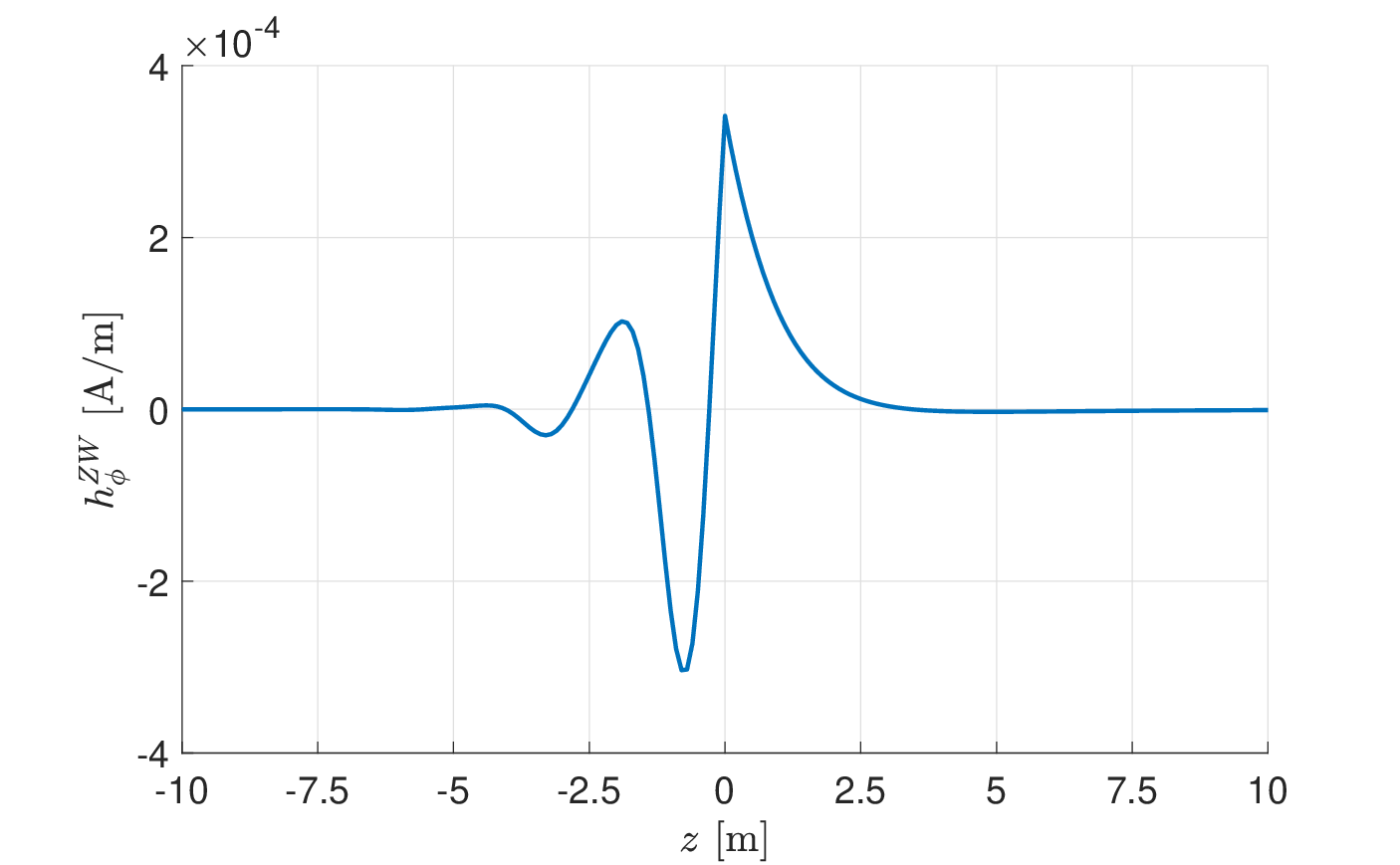}
		\caption{}
		\label{fig:da}
	\end{subfigure}
	\begin{subfigure}{\columnwidth}
		\includegraphics[width=\columnwidth]{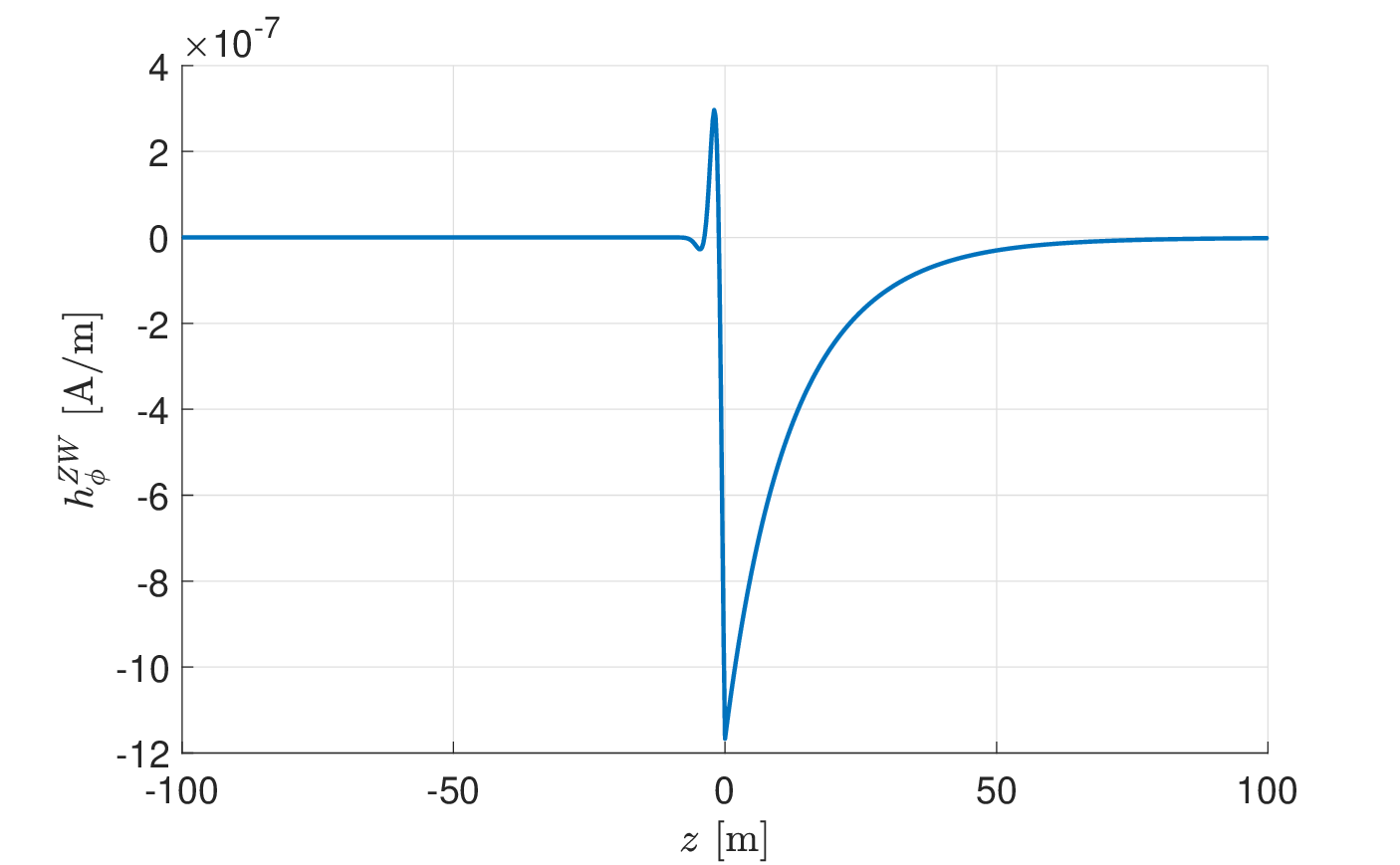}
		\caption{}
		\label{fig:db}
	\end{subfigure}
	\caption{Variation of the modal field $h_{\phi}^{\mathrm{ZW}}$ with the transverse distance $z$ for fixed values of $\rho$. (a) $\rho=5$ m and (b) $\rho=100$ m with $\Delta_0=0.1 \rho$.}
	\label{fig:var_z}
\end{figure}
We examine now the variation of the modal contribution $h_{\phi}^{\mathrm{ZW}}$ with the transverse direction $z$. The $z$-dependent field can be obtained through the formulas \eqref{eqn:2.2.25} and \eqref{eqn:2.2.29} extending them to include the $z$ variation. In fact,  it can easily be shown that it is sufficient to multiply the integrand by the factor
\begin{equation}
	\begin{split}
	&M^{+}(\ko,q;z)= \cos \pt{\kzoq z} \\
	&+\jrm\,\frac{\kzuq}{\ecr(\ko)\,\kzoq}\,
	\sin \pt{\kzoq z}
	\end{split}
\end{equation}
for $z>0$ and
\begin{equation}
		M^{-}(\ko,q;z)= 
		\esp{\jrm \kzuq z}
\end{equation}
for $z<0$. However, when assessing the vertical confinement of a transient contribution, a vertical sweep at \emph{fixed}
$t$ is generally not conclusive. In fact, for a fixed observation point $\rho$, changing $z$
also changes the source--observer distance and consequently the time elapsed since the arrival of the
causal front. We thus define the retarded time (time after arrival)
\begin{equation}
	\Delta(z)=\tau-\sqrt{\rho^2+z^2}.
\end{equation}
As $z$ increases, at fixed $\tau$, the retarded time
$\Delta(z)$ decreases, i.e., the waveform is sampled closer to its arrival. Therefore, confinement in air must be assessed by comparing the field at equal retarded time, i.e.,
by evaluating $h_\phi^{\mathrm{ZW}}(\rho,z,\tau)$ at
\begin{equation}
	\tau(z)=\sqrt{\rho^2+z^2}+\Delta_0,
\end{equation}
with $\Delta_0>0$ fixed. In Fig. \ref{fig:var_z} we report the modal field $h_{\phi}^{\mathrm{ZW}}$ as a function of the transverse distance $z$ for $\rho=5$ m and $\rho=100$ m with $\Delta_0=0.1 \rho$. It can be observed that the field profile is consistent with an evanescent vertical dependence in air, as expected for a SW behavior.
\begin{figure}[t]
	\centering
	\includegraphics[width=\columnwidth]{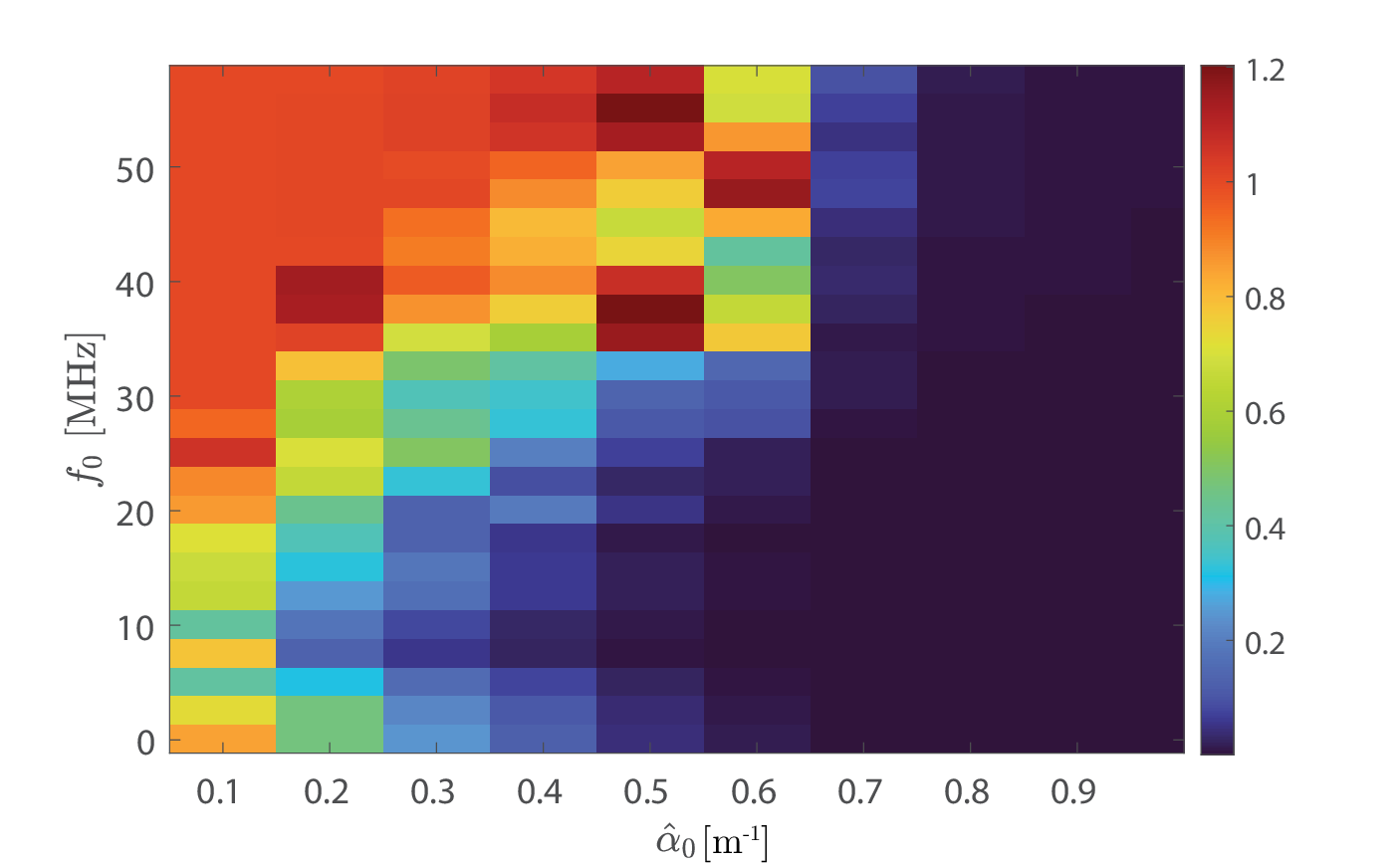}
	\caption{Color map of the dominance parameter $\Gamma$ evaluated at $\rho=100$ m for a lossy ground with $\er=3.2$ and $\sigma=0.02~\mathrm{S/m}$ as a function of the central frequency $f_0$ and of the spectral damping parameter $\hat{\alpha}_0$ of the pulsed excitation in \eqref{eqn:scds}.}
	\label{fig:gamma}
\end{figure}

Finally, we connect these observations to the Zenneck resonance condition \eqref{eqn:spole} in the spectral domain.
Although the Zenneck pole $k_\rho^{\mathrm{ZW}}$ is not enclosed by the $\kr$-plane SDP deformation and therefore does not appear as an explicit term in the FD SDP representation, the dominant TD modal pole generated by the second (frequency-plane) deformation is a true root of $D_0(k_0,q)=0$.
In fact, for the dominant branch $k_0^{\mathrm{ZW}}(q)$, letting
$
\kr^{\mathrm{ZW}}(q)=k_0^{\mathrm{ZW}}(q)-\jrm q
$
and substituting it into \eqref{eqn:spole} yields exactly $D_0(k_0,q)=0$ on the analytically continued sheet selected by the $\mathrm{SDP}_0$ deformation.
Therefore, the mapped pair $\pq{k_0^{\mathrm{ZW}}(q),k_\rho^{\mathrm{ZW}}(q)}$ lies on the Zenneck dispersion manifold throughout the $q$-range.

\subsection{Zenneck-wave late-time dominance}
Having clarified the characteristics of the TD ZW contribution, we now show that, under suitable conditions, it can become dominant over a finite and physically relevant late-time interval. This point must be distinguished from the strict asymptotic behavior for $\tau\to\infty$ at fixed $\rho$, which is discussed separately in Appendix~\ref{app:asymptotic_tail}: the ultimate algebraic tail is of order $\tau^{-5/2}$ and receives contributions not only from the residual continuous spectrum, but also from the modal family generated by $D_0$. Accordingly, the TD ZW contribution discussed here should be regarded as a dominant finite-late-time component, rather than as the whole strict asymptotic tail.

We start by defining a late-time window
$\mathcal{W}(\rho)=[\tau_{\rho,1}(\rho),\tau_{\rho,2}(\rho)]$
automatically selected for each observation point $\rho$ from the corresponding \emph{total} field waveform.
In particular, on the $\tau_\rho$ axis we introduce the smoothed local-RMS envelope
$
s_{\mathrm{ref}}(\tau_\rho;\rho)=
\sqrt{\mathcal{M}\!\left\{|h_{\phi}(\rho,\tau_\rho)|^2\right\}},
$
where $\mathcal{M}\{\cdot\}$ denotes a short moving-average operator.
The window is then started shortly after the main peak of $s_{\mathrm{ref}}(\cdot;\rho)$ and ended at the first $\tau_\rho$ after which $s_{\mathrm{ref}}(\cdot;\rho)$ remains below $2\%$ of its peak for a quiet interval equal to $5\%$ of the total reduced-time observation window.
This provides a robust late-time interval in which the transient at distance $\rho$ has entered its tail regime.

 To quantify whether the ZW contribution alone provides a good approximation of the total late-time field, we introduce the dominance parameter
\begin{equation}
	\Gamma(\rho)=
	\frac{\| h_\phi(\rho,\tau_\rho)-h_\phi^{\mathrm{ZW}}(\rho,\tau_\rho)\|_{\mathcal W}}
	{\|h_\phi(\rho,\tau_\rho)\|_{\mathcal W}},
	\label{eq:Gamma_def}
\end{equation}
where $\|a\|_{\mathcal W}^2=\int_{\mathcal W}|a(\rho,\tau_\rho)|^2\,\mathrm d\tau_\rho$.
Accordingly, a ZW-like \emph{dominance} regime is characterized by $\Gamma(\rho)\ll 1$.

In Fig. \ref{fig:gamma} we thus report the dominance parameter $\Gamma$ at $\rho=100$ m for the nominal case $\er=3.2$ and $\sigma=0.02$ as a function of the central frequency $f_0$ and of the normalized attenuation $\hat{\alpha}_0$ of the pulsed excitation in \eqref{eqn:scds}: small values of $\Gamma$ indicate that the ZW-like modal contribution $h_\phi^{\mathrm{ZW}}$ alone accurately reproduces the total field over the finite late-time window $\mathcal W$.
\begin{figure}[t]
	\centering
	\includegraphics[width=\columnwidth]{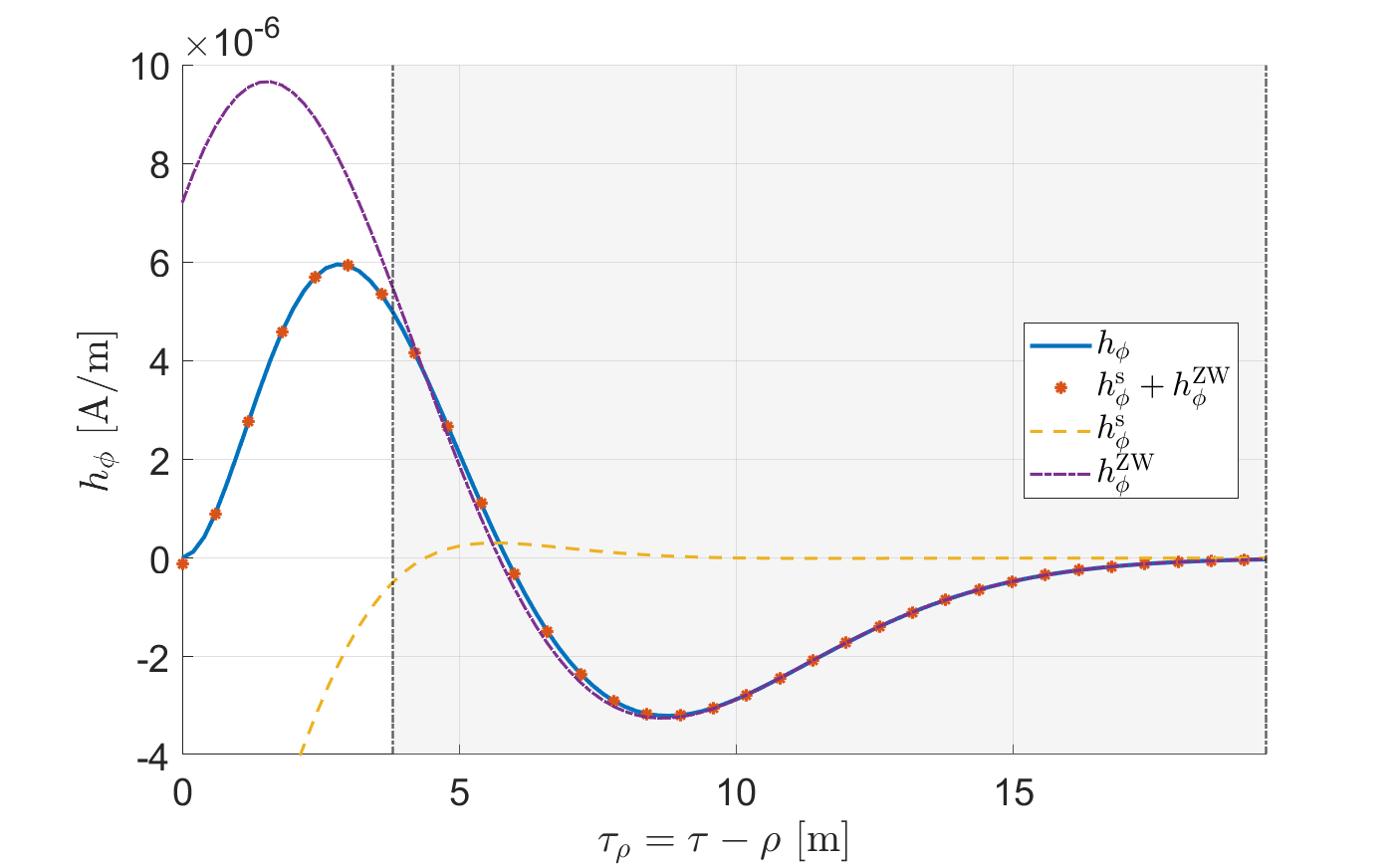}
	\caption{TD field $h_\phi(\rho,t)$ at $\rho=100~\mathrm{m}$ for a pulsed source as in \eqref{eqn:scds} with $f_0=25$ MHz and $\hat{\alpha}_0=0.8~\mathrm{m}^{-1}$: total field $h_{\phi}$, its approximation as $h_{\phi 0}^{\srm} + h_{\phi}^{\mathrm{ZW}}$ and the single components $h_{\phi 0}^{\srm}$ and $h_{\phi}^{\mathrm{ZW}}$.}
	\label{fig:domzw}
\end{figure}  
\begin{figure}[t]
	\centering
	\includegraphics[width=\columnwidth]{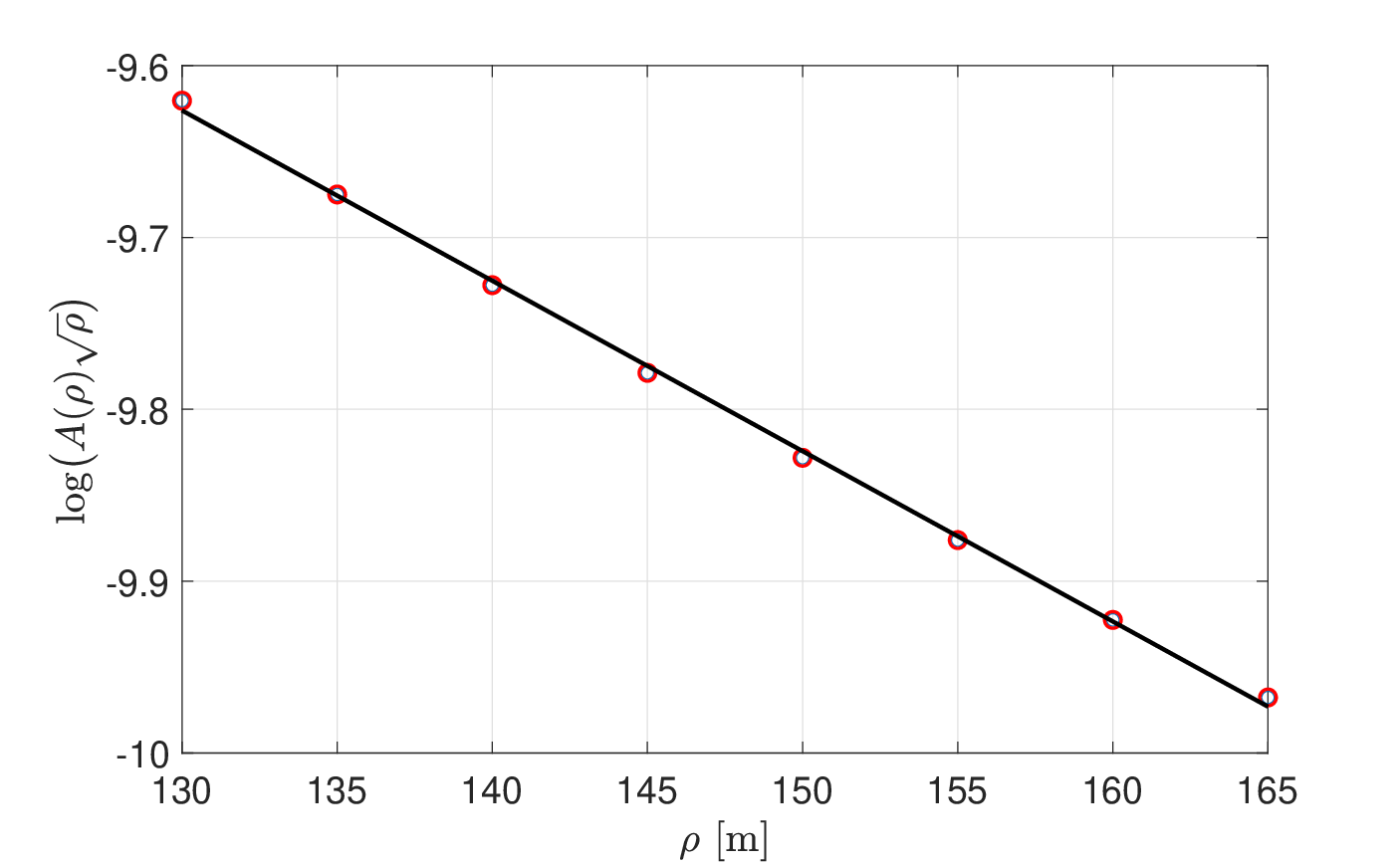}
	\caption{Spatial attenuation of the modal contribution $h_{\phi}^{\mathrm{ZW}}$ for a configuration as in Fig. \ref{fig:domzw}.}
	\label{fig:attenuation_fit3}
\end{figure}
\begin{figure}[t]
	\centering
	\begin{subfigure}{\columnwidth}
		\includegraphics[width=\columnwidth]{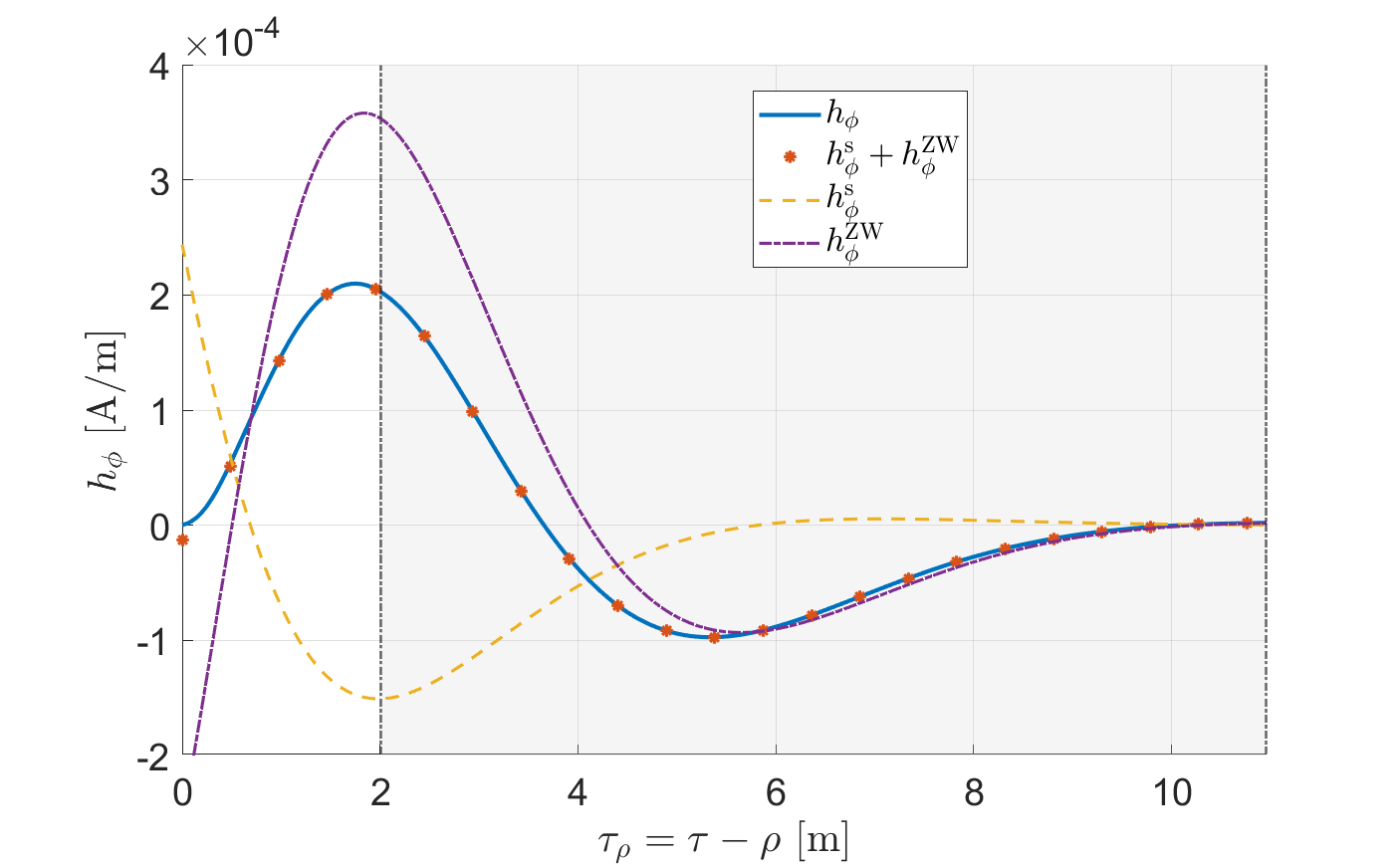}
		\caption{}
		\label{fig:dm1}
	\end{subfigure}
	\begin{subfigure}{\columnwidth}
		\includegraphics[width=\columnwidth]{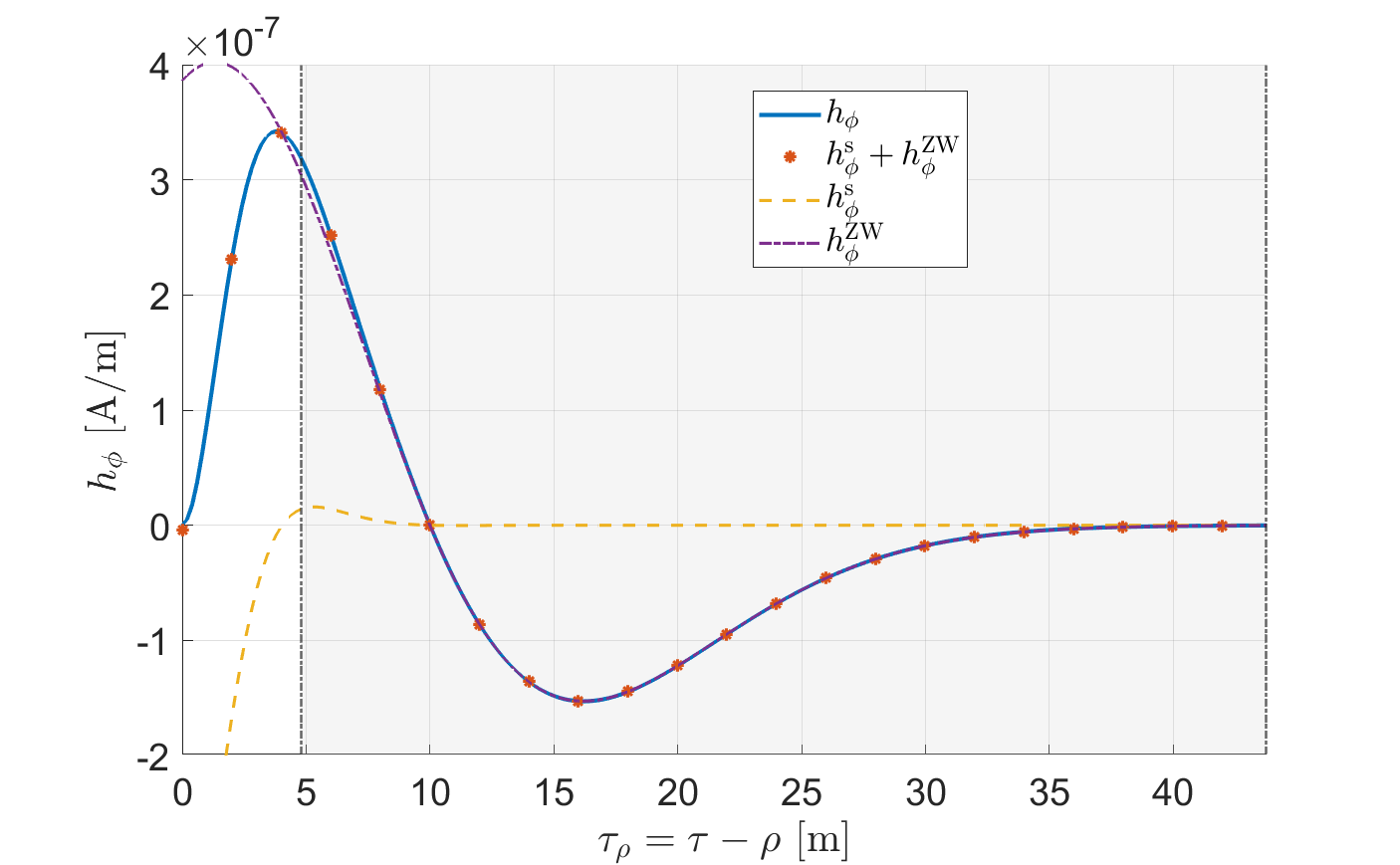}
		\caption{}
		\label{fig:dm2}
	\end{subfigure}
	\caption{Same as in Fig. \ref{fig:domzw}, but with $\rho=10$ m (a) and $\rho=500$ m (b).}
	\label{fig:domzw2}
\end{figure}
The map shows a nearly monotonic improvement as the spectral damping parameter $\hat{\alpha}_0$ increases: for instance, around $f_0\simeq 30~\mathrm{MHz}$ the parameter decreases from $\Gamma \simeq 1$ at weak source damping down to $\Gamma\simeq 2.5\times 10^{-2}$ at strong source damping. In general, a clear ``ZW-dominant'' region (e.g., $\Gamma < 0.1$) appears for sufficiently large values of $\hat{\alpha}_0$, typically for $\hat{\alpha}_0 \gtrsim 0.7~\mathrm{m}^{-1}$ in the range $f_0\in[10,40]~\mathrm{MHz}$. This is consistent with the fact that increasing $\hat{\alpha}_0$ suppresses the slowly decaying source-driven oscillatory behavior, so that the finite late-time response is increasingly dominated by $h_{\phi}^{\mathrm{ZW}}$. For weak source damping ($\hat{\alpha}_0\lesssim 0.3~\mathrm{m}^{-1}$), $\Gamma$ remains close to unity (and can slightly exceed 1 in some cases due to partial cancellation between modal and non-modal contributions in the total field), confirming that $h_\phi^{\mathrm{ZW}}$ alone is insufficient to approximate the late-time waveform in that regime. 
\begin{figure}[t]
	\centering
	\includegraphics[width=\columnwidth]{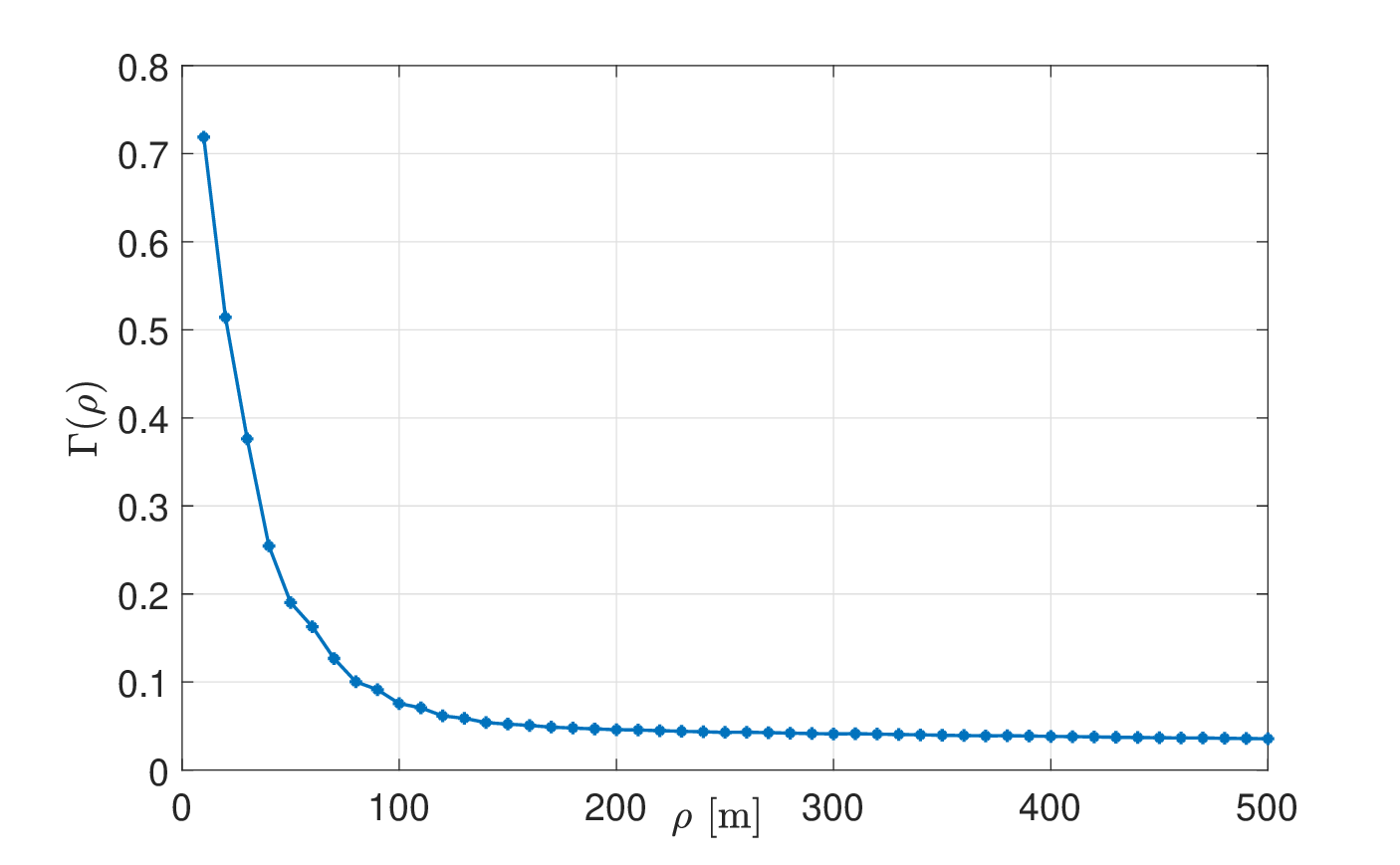}
	\caption{Variation of the dominance parameter $\Gamma$ as a function of the lateral distance $\rho$ for the configuration of Fig. \ref{fig:domzw}.}
	\label{fig:gammar}
\end{figure}  
\begin{figure}[t]
	\centering
	\begin{subfigure}{\columnwidth}
		\includegraphics[width=\columnwidth]{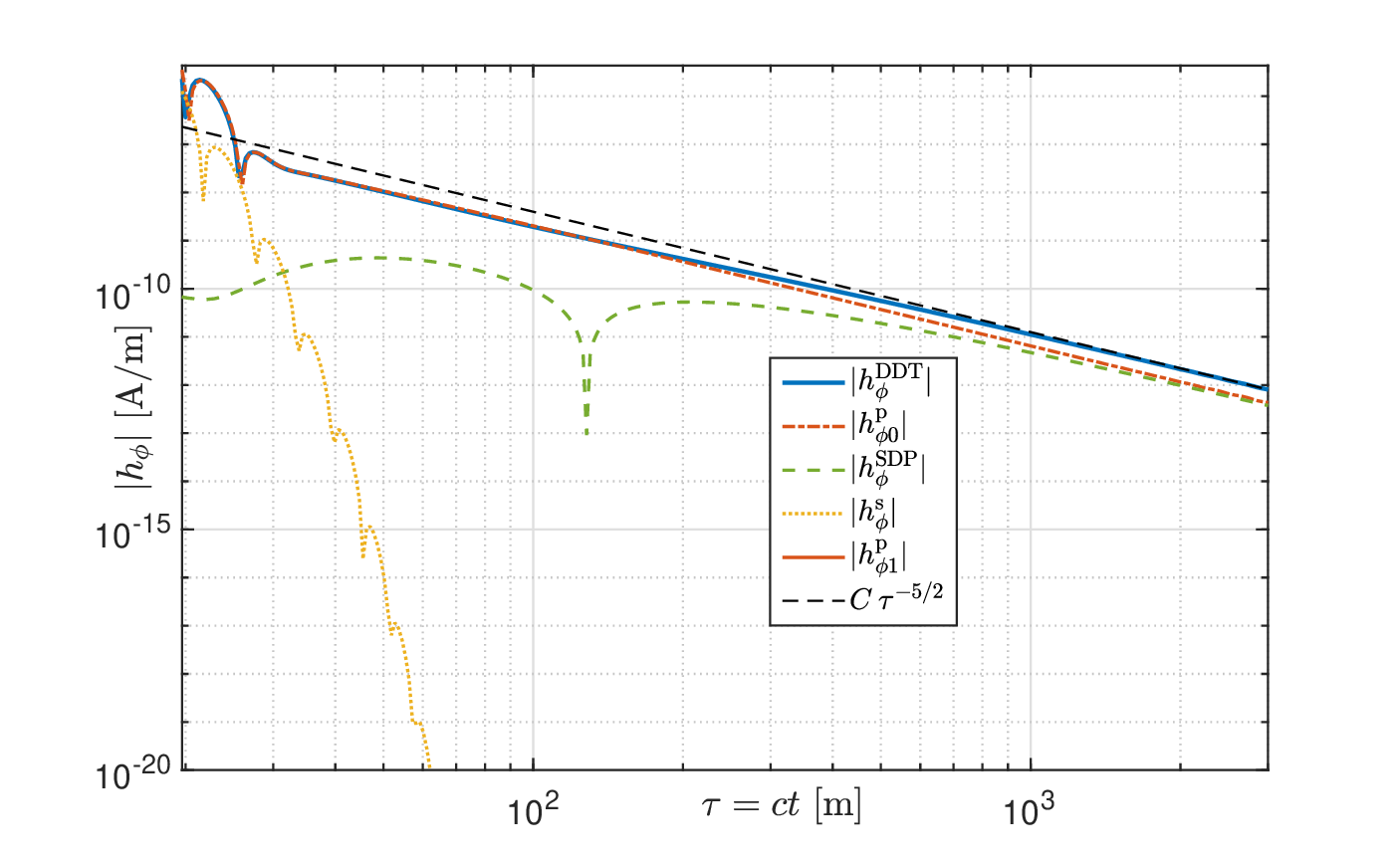}
		\caption{}
		\label{fig:asym1}
	\end{subfigure}
	\begin{subfigure}{\columnwidth}
		\includegraphics[width=\columnwidth]{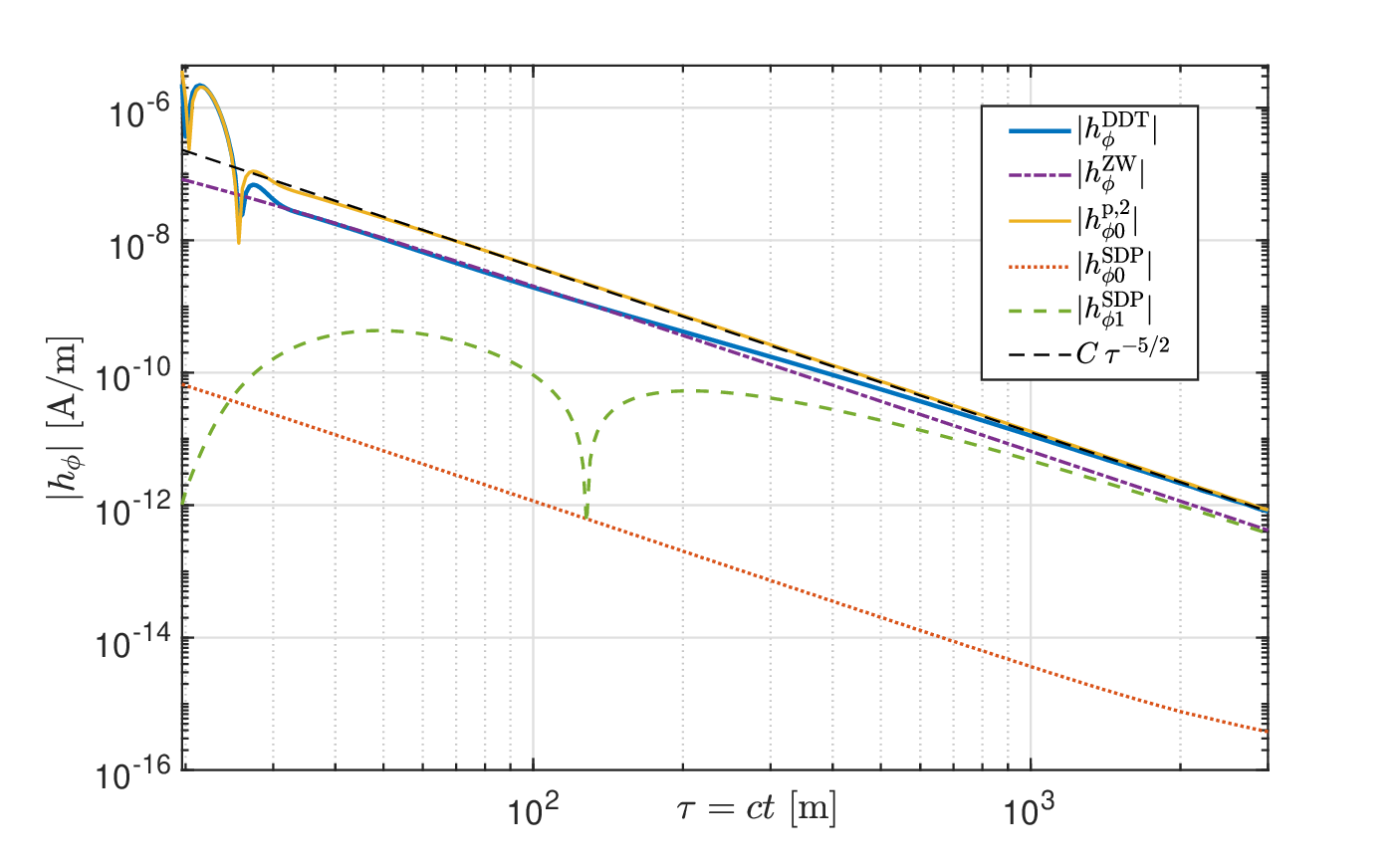}
		\caption{}
		\label{fig:asym2}
	\end{subfigure}
	\caption{Asymptotic tails for the structure in Fig. \ref{fig:domzw2}(a). Total field and different constituents to the DDT representation (a); details of the modal-pole $D_0$-family and continuous-spectrum SDP terms (b).}
	\label{fig:asym}
\end{figure}

As an example, we consider a pulsed source as in \eqref{eqn:scds} with $f_0=25$ MHz  and $\hat{\alpha}_0=0.8~\mathrm{m}^{-1}$. In Fig. 
\ref{fig:domzw}, we report the total field and its approximation \eqref{eq:reduced_model} together with the single contributions $h_{\phi 0}^{\srm}$ and $h_{\phi}^{\mathrm{ZW}}$ for $\rho=100$ m. As it can be seen, the TD ZW is perfectly superimposed to the total field within the selected finite late-time window  (\emph{gray shaded}): in such a case, we have $\Gamma = 0.07$. To confirm the relation between the modal contribution $h_{\phi}^{\mathrm{ZW}}$ and the FD ZW,  the fit of the $y(\rho)$ function is reported in Fig.~\ref{fig:attenuation_fit3} which yields an effective attenuation $\alpha_\rho=0.01\,\mathrm{m^{-1}}$ with a coefficient of determination $R^2=0.99$ while   $\alpha_{\rho}^{\mathrm{ZW}}(\omega_0)=0.016\,\mathrm{m^{-1}}$.

However, as said, the dominance parameter $\Gamma$ depends on the observation point: in Fig. \ref{fig:domzw2} we report the same as in Fig. 
\ref{fig:domzw}, but at $\rho=10$ m (a) for which $\Gamma=0.7$ and $\rho = 500$ m (b) for which $\Gamma =0.03$.  It is thus clear that the physical reality of the TD ZW strongly depends on the characteristics of the pulsed source \emph{and} on the observation point. For completeness, in Fig. \ref{fig:gammar} we report the $\Gamma$ parameter as a function of $\rho$ for the configuration of Fig. \ref{fig:domzw}.
\begin{figure}[t]
	\centering
	\includegraphics[width=\columnwidth]{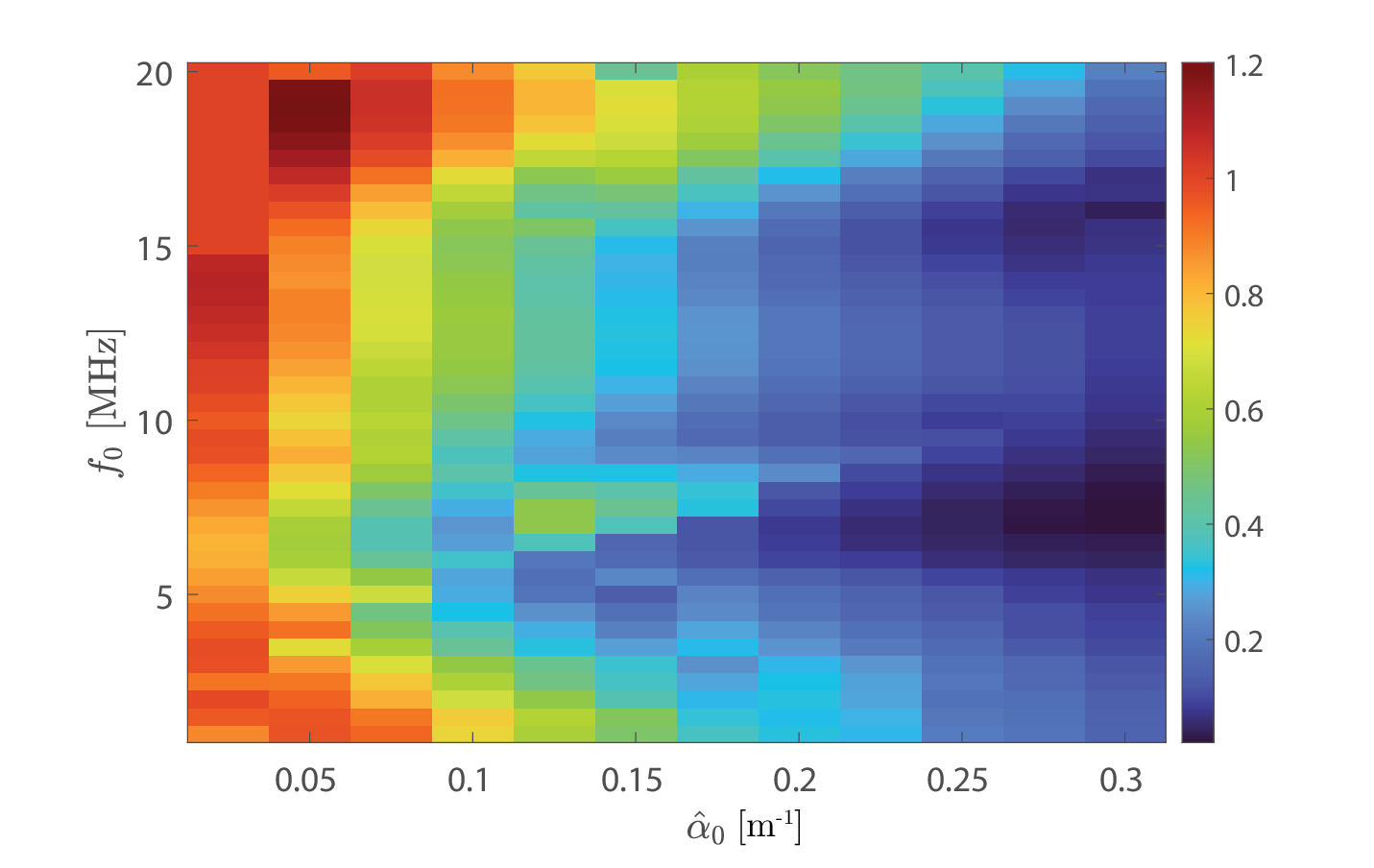}
	\caption{Color map of the dominance parameter $\Gamma$ evaluated at $\rho=1$ km for a lossy ground with $\er=20$ and $\sigma=0.1~\mathrm{S/m}$ as a function of the central frequency $f_0$ and of the spectral damping parameter $\hat{\alpha}_0$ of the pulsed excitation in \eqref{eqn:scds}.}
	\label{fig:gammaclay}
\end{figure}
\begin{figure}[t]
	\centering
		\includegraphics[width=\columnwidth]{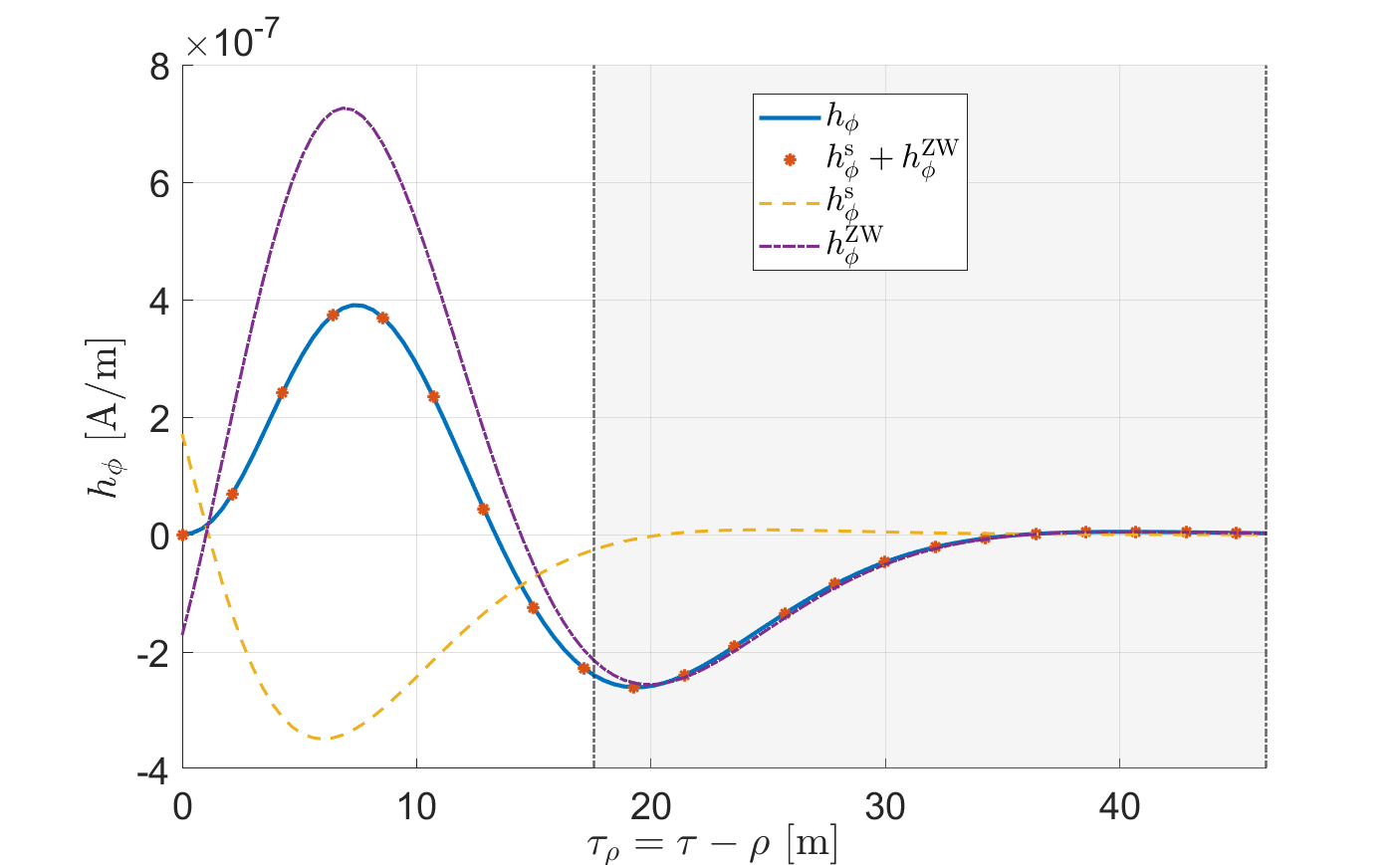}
	\caption{TD field $h_\phi(\rho,t)$ at $\rho=1~\mathrm{km}$ for a structure as in Fig. \ref{fig:gammaclay} with $f_0=7$ MHz and $\hat{\alpha}_0=0.25~\mathrm{m}^{-1}$: total field $h_{\phi}$, its approximation as $h_{\phi 0}^{\srm} + h_{\phi}^{\mathrm{ZW}}$ and the single components $h_{\phi 0}^{\srm}$ and $h_{\phi}^{\mathrm{ZW}}$.}
	\label{fig:dom2}
\end{figure}

Finally, in Fig. \ref{fig:asym} the asymptotic tails of the field are reported for the case considered in Fig. \ref{fig:dm1}. In particular, in Fig. \ref{fig:asym}(a) the DDT total field $h_{\phi}^{\mathrm{DDT}}$ with its basic constituents $ h_{\phi}^{\srm}$, $h_{\phi 0}^{\prm}$, $h_{\phi 1}^{\prm}$,
$h_{\phi}^{\mathrm{SDP}}=h_{\phi 0}^{\mathrm{SDP}}+h_{\phi1}^{\mathrm{SDP}}$, showing an excellent agreement with the theoretical asymptotic decays (as predicted in Appendix). In particular, the source-pole $ h_{\phi}^{\srm}$  and the modal-pole $D_1$-family $h_{\phi 1}^{\prm}$ terms decay exponentially (the latter is not visible since it is off scale), while both the modal-pole $D_0$-family $h_{\phi 0}^{\prm}$ and  the continuous-spectrum SDP contribution $h_{\phi}^{\mathrm{SDP}}$ decay as $\tau^{-5/2}$.  Interestingly, as shown in Fig. \ref{fig:asym}(b), in the asymptotic regime, in addition to the TD-ZW term $h_{\phi}^{\mathrm{ZW}}$ also the modal-pole term $h_{\phi 0}^{\prm, 2}$ (i.e., the purely imaginary pole) and the continuous-spectrum SDP contribution $h_{\phi 1}^{\mathrm{SDP}}$ contribute to the total field with a $\tau^{-5/2}$ decay while all the other contributions remains negligible in all the temporal range. In particular, we observe a transition from a total field dominated by the $h_{\phi}^{\mathrm{ZW}}$ term to a field represented by the $h_{\phi 0}^{\prm, 2}$ (where the field, however, has decayed by more than five orders of magnitude).

The results discussed so far refer to the nominal configuration $\er=3.2$ and $\sigma=0.02~\mathrm{S/m}$, but the presented results hold also for a more general configuration with arbitrary values of $\er$ and $\sigma$. As a further example, we consider a lossy ground with $\er=20$ and $\sigma=0.1~\mathrm{S/m}$ (wet clay) and in Fig. \ref{fig:gammaclay} we thus report the dominance parameter $\Gamma$ at $\rho=1$ km as a function of the central frequency $f_0$ and of the normalized attenuation $\hat{\alpha}_0$ of the pulsed excitation. As expected, $\Gamma$ decreases by increasing $\hat{\alpha}_0$.

\begin{figure}[t]
	\centering
	\includegraphics[width=\columnwidth]{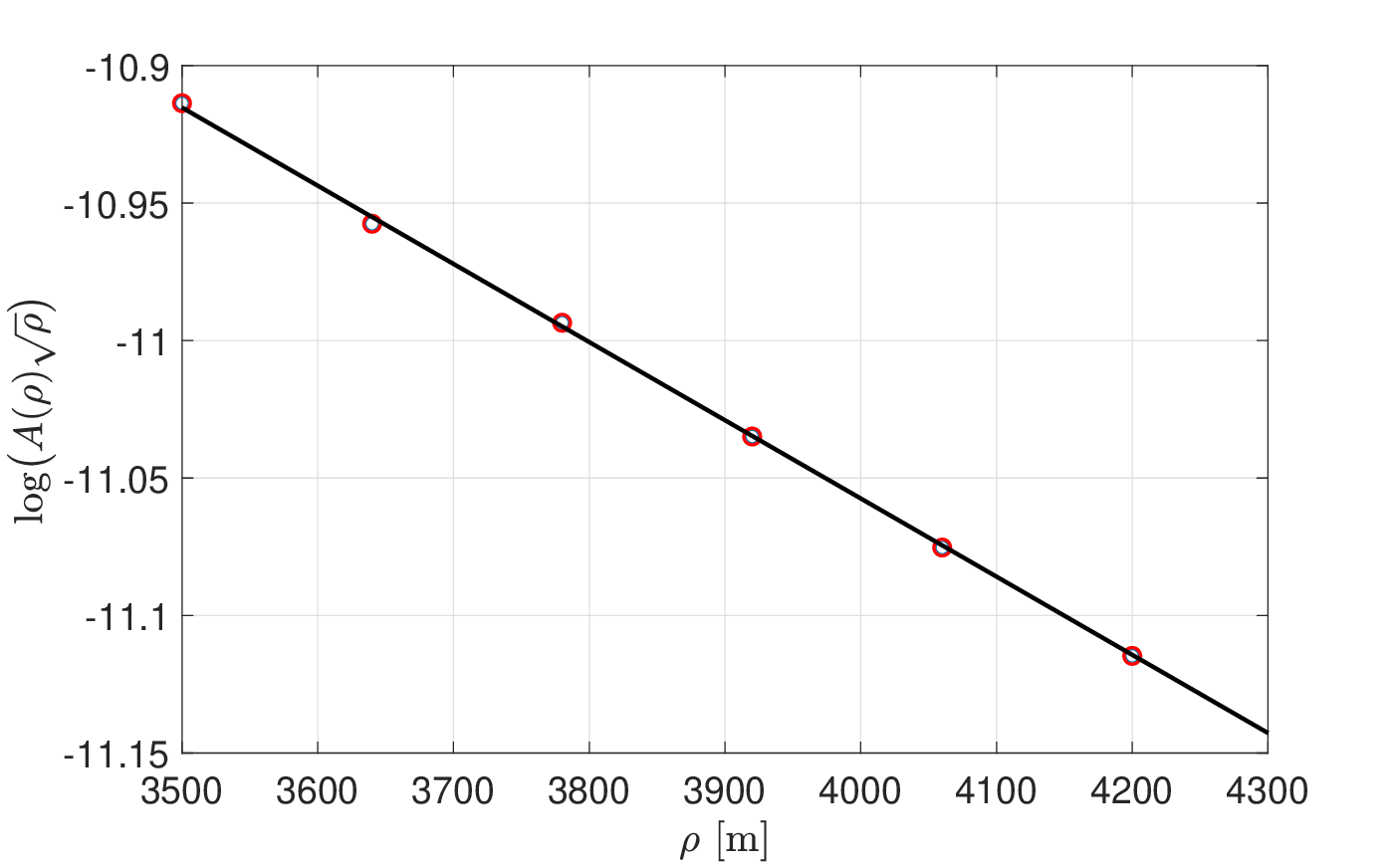}
	\caption{Spatial attenuation of the modal contribution $h_{\phi}^{\mathrm{ZW}}$ for a configuration as in Figs. \ref{fig:gammaclay}-\ref{fig:dom2}}
	\label{fig:attenuation_fit2}
\end{figure}

In Fig. \ref{fig:dom2}, we report the total field and its approximation \eqref{eq:reduced_model} together with the separate contributions $h_{\phi 0}^{\srm}$ and $h_{\phi}^{\mathrm{ZW}}$ for $\rho=1$ km with $f_0=7$ MHz and $\hat{\alpha}_0=0.25~\mathrm{m}^{-1}$ ($\Gamma=0.04$): the results confirm the dominance of the TD ZW in the late time regime (gray shaded).  Finally, in Fig.~\ref{fig:attenuation_fit2} we report the function $y(\rho)=\ln\!\pq{A(\rho)\sqrt{\rho}}$ over a set of observation points. This yields an effective attenuation $\alpha_\rho=2.8 \cdot 10^{-4}\,\mathrm{m^{-1}}$ with a coefficient of determination $R^2=0.99$ and, in this case, it coincides with    $\alpha_{\rho}^{\mathrm{ZW}}(\omega_0)$ thus confirming the ZW-like behavior of the modal contribution $h_{\phi}^{\mathrm{ZW}}$.

Therefore, the numerical results show that the proposed DDT isolates a dominant modal contribution that carries the TD footprint of the ZW and, under suitable conditions, can govern a broad and physically relevant finite late-time interval.

\section{Conclusions}
This work presented a rigorous and causal time-domain analysis of the field radiated by a pulsed vertical electric dipole above a lossy half-space. By adapting the DDT to the canonical Sommerfeld half-space problem, we derived an explicit separation of the total field into source-pole, loss-pole, modal-pole, and steepest-descent contributions.

The proposed DDT solution is first validated through comparisons with an accurate direct double inverse transform, showing full agreement of the resulting time-domain waveforms. More importantly, the decomposition reveals that the time-domain field can be accurately represented, over broad and physically relevant finite late-time intervals, by the sum of a single dominant modal-pole contribution and a source-pole contribution generated by the frequency-plane deformation. The dominant modal term exhibits clear surface-wave-like footprints, namely near-invariance of the waveform when expressed in reduced time $\tau_\rho=\tau-\rho$ and an attenuation trend consistent with $\exp(-\alpha_\rho \rho)/\sqrt{\rho}$. Under suitable excitation and observation conditions, this term can dominate the transient response over a broad finite late-time interval. At the same time, for the considered damped-sinusoidal excitation, the strict asymptotic tail for $t\to\infty$ at fixed $\rho$ is algebraic of order $t^{-5/2}$ and in general it receives contributions from both the continuous spectrum and different modal-pole terms. Although the Zenneck pole is not enclosed by the transverse-wavenumber steepest-descent deformation and does not appear as an explicit term in the first step, its pole physics re-enters through the frequency-plane deformation via these time-domain modal contributions, yielding a tangible and interpretable signature of the conventional frequency-domain Zenneck wave.

Work is in progress to extend the formulation to pulses of finite duration and to noble metals at optical wavelengths where the Zenneck wave turns into a plasmon mode.

\appendices
\section{}
\label{app:asymptotic_tail}

For \(\tau>\rho\sqrt{\er}\), the field admits the decomposition
\begin{equation}
	\begin{split}
		h_{\phi}(\rho,\tau)&=
		h_{\phi0}^{\srm}(\rho,\tau)+
		h_{\phi1}^{\srm}(\rho,\tau)+
		h_{\phi0}^{\prm}(\rho,\tau)\\
		&+
		h_{\phi1}^{\prm}(\rho,\tau)+
		h_{\phi0}^{\mathrm{SDP}}(\rho,\tau)+
		h_{\phi1}^{\mathrm{SDP}}(\rho,\tau) .
	\end{split}
\end{equation}
In this appendix we determine which of these terms contribute to the \emph{strict} late-time tail at fixed \(\rho\). We show that the source-pole terms are exponentially small, that the modal family generated by \(D_1\) is also exponentially small, and that the algebraic tail of order \(\tau^{-5/2}\) arises from the modal terms associated with \(D_0\) together with the residual SDP contributions.

We assume throughout a causal physically realizable excitation, so that \(\hat I(k_0)\) is analytic in the lower half-plane and regular at \(k_0=0\). For the damped-sinusoidal source used in Sec.~VI one also has \(\hat I(0)\neq 0\).

The source-pole terms are exponentially small. In fact, if \(\hat I(k_0)\) has an $n$-th order pole at \(k_0=k_0^{\srm}\) with \(\im{k_0^{\srm}}>0\),
\begin{equation}
	h_{\phi i}^{\srm}(\rho,\tau)
	=
	\re{
		\esp{\jrm k_0^{\srm}\tau}
		\sum_{m=0}^{n-1} a_{i,m}(\rho)\,\tau^m
	},
	\qquad i=0,1,
\end{equation}
for suitable \(\tau\)-independent coefficients \(a_{i,m}(\rho)\). Therefore
\begin{equation}
	h_{\phi i}^{\srm}(\rho,\tau)
	=
	O \bigl(\tau^n \esp{-\im{k_0^{\srm}}\tau}\bigr),
	\qquad \tau\to+\infty .
\end{equation}
For the damped-sinusoidal excitation of Sec.~VI, this gives
\begin{equation}
	h_{\phi0}^{\srm},\ h_{\phi1}^{\srm}
	=
	O \bigl(\tau \esp{-\hat\alpha_0\tau}\bigr).
\end{equation}

We next consider the modal contribution generated by \(D_0(k_0,q)\). For \(q\to0^+\), the three roots behave as
\begin{align}
	k_{0\pm}(q)
	&=
	\pm \sqrt{2\ks q}
	+\jrm \frac{4\er+3}{4}\,q
	+O(q^{3/2}),
	\\
	k_{03}(q)
	&=
	\jrm \frac{q}{2}+O(q^2).
\end{align}
Evaluation of the corresponding residues shows that the complex pole contributes with amplitude \(q^{1/4}\), whereas the purely imaginary branch contributes with amplitude \(q^{3/2}\). Since
\(
\esp{\jrm k_{0\pm}(q)\tau}
\)
contains the oscillatory factor
\(
\esp{\pm \jrm \sqrt{2\ks q}\,\tau},
\)
the endpoint \(q\to0^+\) determines the late-time behavior; with the change of variable \(q=s^2\), all these modal terms yield the same decay order and, in particular,
\begin{equation}
	h_{\phi0}^{\prm}(\rho,\tau)
	\sim
	C_{\prm,0}(\rho)\,\tau^{-5/2},
	\qquad \tau\to+\infty ,
\end{equation}
for a suitable coefficient \(C_{\prm,0}(\rho)\).

For the modal terms associated with \(D_1(k_0,q)\), the branches originating from the double root at \(u=\ks/\er\) satisfy
\begin{equation}
	k_0(q)=\jrm\frac{\ks}{\er}+\jrm\frac{q^2}{4\ks}+O(q^3),
	\qquad q\to0^+,
\end{equation}
so that the corresponding temporal factor is
\begin{equation}
	\esp{\jrm k_0(q)\tau}
	=
	\esp{-\frac{\ks}{\er}\tau}
	\esp{-\frac{q^2}{4\ks}\tau+o(q^2\tau)}.
\end{equation}
Therefore the associated contribution is exponentially small.

We now turn to the residual double-integral SDP terms. For \(h_{\phi0}^{\mathrm{SDP}}\), after setting \(k_0=\jrm u\), the relevant inner \(q\)-integral is supported on \(0\le q\le 2u\). The scaling \(q=us\), \(0\le s\le2\), gives an apparent \(O(u)\) contribution, but its coefficient vanishes identically, i.e.,
\begin{equation}
	\int_0^2 \frac{1-s}{\sqrt{s(2-s)}}\,\dd s =0.
\end{equation}
The dominant contribution therefore comes from the endpoint regions \(q=O(u^2)\) and \(2u-q=O(u^2)\), both of which produce \(O(u^{3/2})\). As a result,
\begin{equation}
	h_{\phi0}^{\mathrm{SDP}}(\rho,\tau)
	\sim
	C_{\mathrm{SDP},0}(\rho)\,\tau^{-5/2},
	\qquad \tau\to+\infty .
\end{equation}

Finally, we consider \(h_{\phi1}^{\mathrm{SDP}}\). Writing
\begin{equation}
	\kappa=\sqrt{\er u^2-\ks u}
	=
	\jrm\sqrt{\ks u}+O(u^{3/2}),
	\qquad u\to0^+,
\end{equation}
one finds that the relevant inner scaling is
$
q=O(\sqrt{u}),
$
which yields an \(O(u^{3/2})\) contribution to the inner \(q\)-integral. In the outer region \(q\gg\sqrt{u}\), the integrand has the expansion
\begin{equation}
	-\frac{u}{\ks}\,q\,\bh_1^{(2)}(-\jrm q\rho)+O(u^{3/2}),
\end{equation}
and the leading \(O(u)\) outer coefficient is real. Since \(h_{\phi1}^{\mathrm{SDP}}\) is obtained by taking the real part of an expression multiplied by the prefactor \(\jrm\), that \(O(u)\) term does not contribute to the physical field. Therefore the first nonvanishing algebraic contribution is again of order \(u^{3/2}\), so that
\begin{equation}
	h_{\phi1}^{\mathrm{SDP}}(\rho,\tau)
	=
	O\!\pt{\tau^{-5/2}},
	\qquad \tau\to+\infty .
\end{equation}

Collecting all the previous estimates, one finally obtains
\begin{equation}
	h_{\phi}(\rho,\tau)
	=
	O\!\pt{\tau^{-5/2}},
	\qquad \tau\to+\infty
\end{equation}
at fixed \(\rho\). Therefore the  late-time tail is algebraic of order \(\tau^{-5/2}\), and it receives contributions from both the SDP continuous spectrum and the modal-pole terms, including \(h_{\phi}^{\mathrm{ZW}}\).

\bibliographystyle{ieeetr}
\bibliography{IEEEabrv,Bib_Transient_Zenneck}
	
\end{document}